\documentclass[longauth]{aa}

\usepackage{graphicx}
\usepackage{txfonts}
\usepackage[T1]{fontenc}

\usepackage{subcaption} 
\usepackage{color}
\usepackage{amssymb}
\usepackage{epstopdf}
\usepackage{natbib}
\bibpunct{(}{)}{;}{a}{}{,} 
\usepackage{url}

\newcommand{\vsini}{$V \sin i_\star$}   
 
\newcommand{\vmic}{$V_{\rm mic}$}
\newcommand{\vmac}{$V_{\rm mac}$}

\newcommand{\teff}{$T_{\rm eff}$}
\newcommand{\logg}{log\,({\it g$_\star$})}

\newcommand{\feh}{[Fe/H]}

\newcommand{\kms}{km\,s$^{-1}$}

\newcommand{\gc}{g~cm$^{-3}$}

\newcommand{\Lsun}{$L_{\odot}$}                          
\newcommand{\Msun}{$M_{\odot}$}
\newcommand{\Rsun}{$R_{\odot}$}

\newcommand{\rjup}{$R_\mathrm{J}$}
\newcommand{\mjup}{$M_\mathrm{J}$}

\newcommand{\Prot}{$P_{rot}$} 
\newcommand{\mstar}{$M_\star$}
\newcommand{\rstar}{$R_\star$}

\newcommand{\rplanet}{$R_{\mathrm{BD}}$}

\newcommand{\targetaa}{EPIC\,212036875}

\newcommand{\bjdtdb}{\ensuremath{\rm {BJD_{TDB}}}}
 
\newcommand{\smasst}[1][]{$1.19 \pm 0.09$}   
\newcommand{\smassp}[1][]{$1.10 \pm 0.04$}   
\newcommand{\smasssouth}[1][]{$1.15 \pm 0.08$}   
\newcommand{\smassspectral}[1][]{$1.21$}   
\newcommand{\smepic}[1][]{$1.21\pm 0.11$}  
\newcommand{\sradiust}[1][]{$1.43\pm 0.29$}  
\newcommand{\sradiusp}[1][]{$1.52\pm 0.06$}  
\newcommand{\sradiussouth}[1][]{$1.41 \pm 0.05$}   
\newcommand{\sradiusspectral}[1][]{$1.30$}   
\newcommand{\sradiusgaia}[1][]{$1.54\pm 0.10$}  
\newcommand{\srgaia}[1][]{$1.49\pm 0.05$}  
\newcommand{\srepic}[1][]{$1.38^{+0.32}_{-0.17}$}  
\newcommand{\steffsme}[1][]{$6230 \pm 90$} 
\newcommand{\steffgaia}[1][]{$6227 \pm 100$} 
\newcommand{\steffepic}[1][]{$6336$} 
\newcommand{\steffkea}[1][]{$6380 \pm 58$} 
\newcommand{\sloggCasme}{$4.20 \pm 0.20$} 
\newcommand{\sloggMgsme}[1][]{$4.17 \pm 0.10$} 
\newcommand{\scahsme}[1][]{$-0.14 \pm 0.05$} 
\newcommand{\sfehsme}[1][]{$-0.28 \pm 0.05$} 
\newcommand{\snasme}[1][]{$-0.11 \pm 0.05$} 
\newcommand{\smghsme}[1][]{$-0.16 \pm 0.05$} 
\newcommand{\sfehepic}[1][]{$-0.08$} 
\newcommand{\sloggp}[1][]{$4.10 \pm 0.04$} 
\newcommand{\sloggepic}[1][]{$4.21$} 
\newcommand{\svsini}[1][]{$10.8 \pm 1.5$} 
\newcommand{\svmic}[1][]{$1.3$} 
\newcommand{\svmac}[1][]{$5.2$} 
\newcommand {\lstargaia}[1][]{$3.2\pm 0.4$} 
\newcommand {\lstargaiaarchive}[1][]{$3.01^{+0.05}_{-0.07}$} 
\newcommand{\denspyaneti}[1][]{$0.55 \pm 0.04$}  
\newcommand{\densMR}{$0.58 \pm 0.08$}  
\newcommand{\stype}{F7\,V} 
\newcommand{\age}{$5.1\pm 0.9$} 
\newcommand{\prot}{$7.2 \pm 0.5$} 
\newcommand{\parallaxgaia}{$3.238\pm0.048$} 
\newcommand{\velgaia}[1][$\mathrm{km\,s^{-1}}$]{ $-22.7 \pm 1.7$} 
\newcommand{\velsme}[1][$\mathrm{km\,s^{-1}}$]{ $xxx \pm x$} 
\newcommand{\istar}{66}  

\newcommand{\pmra}{$-2.62 \pm 0.08$} 
\newcommand{\pmdec}{$-29.70 \pm 0.05$} 
 
\newcommand{\Tzerob}[1][]{$8098.6791 \pm 0.0002$}  
\newcommand{\Pb}[1][]{$5.16992  \pm 0.00002$}  
\newcommand{\eb}[1][]{$0.134 \pm 0.002$}      
\newcommand{\wb}[1][]{$163 \pm 1$}    
\newcommand{\bb}[1][]{$0.920 _{-0.006} ^{+0.005}$}   
\newcommand{\arb}[1][]{$9.2  \pm  0.2$}   
\newcommand{\rrb}[1][]{$0.0608 \pm 0.0009$} 
\newcommand{\kb}[1][]{$5.289 \pm 0.013$} 
\newcommand{\mpb}{$51  \pm 2$}   
\newcommand{\rpb}{$0.83  \pm 0.03$}   
\newcommand{\ib}[1][]{$83.9 \pm 0.2$}  
\newcommand{\ab}[1][]{$0.060 \pm 0.003$} 
\newcommand{\Fequib}[1][]{$740  \pm 50$}    
\newcommand{\denpb}[1][]{$108_{-13} ^{+15}$} 
\newcommand{\pgrav}[1][]{$5.23 \pm 0.02$}  
\newcommand{\Tequib}[1][]{$1450 \pm 30$}    
\newcommand{\ttotb}[1][]{$2.17 \pm 0.01$} 
\newcommand{\tfullb}[1][]{$0.76 \pm 0.09$} 
\newcommand{\qone}[1][]{ $0.41 \pm 0.10$}    
\newcommand{\qtwo}[1][]{ $0.26  \pm 0.10$}     
\newcommand{\uone}[1][]{ $0.33  \pm 0.14$}    
\newcommand{\utwo}[1][]{ $0.30  \pm 0.13$}   

\newcommand{\fone}{$-21.26 \pm 0.03$} 
\newcommand{\ftwo}{$-21.33 \pm 0.16$} 
\newcommand{\fthree}{$-21.30 \pm 0.01$} 

\newcommand{\gammaone}{$0.064 _{-0.030} ^{+0.046}$} 
\newcommand{\gammatwo}{$0.202 _{-0.129} ^{+0.498}$} 
\newcommand{\gammathree}{$0.0092 _{-0.006} ^{+0.010}$} 
\newcommand{\trjitter}{$0.000025 \pm 0.000005$} 

\begin{document} 
 
 \titlerunning{Greening of the Brown Dwarf Desert}
   \title{Greening of the Brown Dwarf Desert\thanks{This work is done under the framework of the KESPRINT
colla-boration ({\tt{http://kesprint.science}}). KESPRINT is an international consortium devoted to the 
characterisation and research of exoplanets discovered with space-based missions.}}
   
   \subtitle{\targetaa b -- a 51~\mjup~object in a 5 day orbit around an F7\,V star}

\author{Carina~M.~Persson\inst{1}   
\and
Szil\'ard~Csizmadia\inst{2}
\and
Alexander~J.~Mustill\inst{3}   
\and 
Malcolm~Fridlund\inst{1,4} 
\and
Artie~P. Hatzes\inst{5}
\and
Grzegorz~Nowak\inst{6,7}   
\and
Iskra~Georgieva\inst{1} 
\and
Davide~Gandolfi \inst{8}   
\and
Melvyn~B.~Davies\inst{3}
\and
John~H.~Livingston\inst{9} 
\and
Enric~Palle\inst{6,7}   
\and
Pilar~Monta\~nes Rodr\'iguez\inst{6,7}
\and
Michael~Endl\inst{10}  
\and
Teruyuki~Hirano \inst{11}  
 \and
Jorge~Prieto-Arranz\inst{6,7}  
\and 
Judith~Korth\inst{12}
 \and 
Sascha~Grziwa \inst{12}  
\and
Massimiliano~Esposito\inst{5}
\and
Simon~Albrecht\inst{13}  
\and
Marshall~C.~Johnson\inst{14}  
\and
Oscar~Barrag\'an\inst{8,15}  
\and
Hannu~Parviainen\inst{6,7} 
\and
Vincent~Van~Eylen\inst{16} 
\and 
Roi~Alonso~Sobrino\inst{6,7}
\and
Paul~G.~Beck\inst{6,7,17}
\and
Juan~Cabrera\inst{2}
\and
Ilaria~Carleo\inst{18}  
\and
William~D.~Cochran\inst{10}  
\and
Fei~Dai\inst{16,19}  
 \and 
Hans~J.~Deeg\inst{6,7}
\and
Jerome~P.~de~Leon\inst{9}
\and
Philipp~Eigm\"uller\inst{2}
 \and
Anders~Erikson\inst{2}
\and
Akai~Fukui\inst{20}
\and
Luc\'ia~Gonz\'alez-Cuesta\inst{6,7}
\and
Eike~W.~Guenther\inst{5}
\and
Diego~Hidalgo\inst{6,7}
\and
Maria~Hjorth\inst{13}
\and
Petr~Kabath\inst{21}
\and
Emil~Knudstrup\inst{13}
\and  
Nobuhiko~Kusakabe\inst{20,22}  
\and
Kristine~W.~F.~Lam\inst{23}  
\and
Mikkel~N.~Lund\inst{13}
\and
Rafael~Luque\inst{6,7}
\and
Savita~Mathur\inst{6,7}
\and
Felipe~Murgas\inst{6,7}  
\and
Norio~Narita\inst{6,9,20,22,24}  
\and
David~Nespral\inst{6,7}
\and
Prajwal~Niraula\inst{25}  
\and
A.~O.~Henrik~Olofsson\inst{1} 
 \and
Martin~P\"atzold\inst{12}
 \and
Heike~Rauer\inst{2,23}
\and
Seth~Redfield\inst{18}  
\and
Ignasi~Ribas\inst{26,27}  
\and
Marek~Skarka\inst{21,28}  
\and
Alexis~M.~S.~Smith\inst{2}  
\and
Jan~Subjak\inst{21,29}  
\and
Motohide~Tamura\inst{9,20,22}   
 }  
 
    \offprints{carina.persson@chalmers.se}
   \institute{Chalmers University of Technology, Department of Space, Earth and Environment, Onsala Space Observatory,  SE-439 92 Onsala, Sweden.  
    \email{\url{carina.persson@chalmers.se}}
    \and 
Institute of Planetary Research, German Aerospace Center (DLR), Rutherfordstrasse
2, D-12489 Berlin, Germany 
  \and
Lund Observatory, Department of Astronomy \& Theoretical Physics, Lund University, Box 43, SE-221 00 Lund, Sweden 
\and
Leiden Observatory, University of Leiden, PO Box 9513, 2300 RA, Leiden, The Netherlands 
\and
Th\"uringer Landessternwarte Tautenburg,  D-07778 Tautenburg, Germany 
\and
Instituto de Astrof\'isica de Canarias (IAC), 38205 La Laguna, Tenerife, Spain 
\and
Departamento de Astrof\'isica, Universidad de La Laguna, 38206 La Laguna, Tenerife, Spain 
\and
Dipartimento di Fisica, Universit\`a di Torino, via Pietro Giuria 1, I-10125, Torino, Italy 
\and
Department of Astronomy, The University of Tokyo, 7-3-1 Hongo, Bunkyo-ku, Tokyo 113-0033, Japan 
\and
Department of Astronomy and McDonald Observatory, University of Texas at Austin, 2515 Speedway, Austin, TX 78712, USA 
\and
Department of Earth and Planetary Sciences, Tokyo Institute of Technology, 2-12-1 Ookayama, Meguro-ku, Tokio, Japan 
\and
Rheinisches Institut f\"ur Umweltforschung an der Universit\"at zu K\"oln, Aachener Strasse 209, 50931 K\"oln, Germany 
\and
Stellar Astrophysics Centre, Department of Physics and Astronomy, Aarhus University, Ny Munkegade 120, DK-8000 Aarhus C 
\and 
Department of Astronomy, The Ohio State University, 140 West 18th Ave., Columbus, OH 43210 USA  
\and
Sub-department of Astrophysics, Department of Physics, University of Oxford, Oxford OX1 3RH, UK  
\and
Department of Astrophysical Sciences, Princeton University, 4 Ivy Lane, Princeton, NJ, 08544, USA 
\and
Institute f\"ur Physik, Geophysik, Astrophysik und Meteorologie, Karl-Franzens Universit\"at Graz, Univ.-Platz 5, 8010 Graz, Austria
\and
Astronomy Department and Van Vleck Observatory, Wesleyan University, Middletown, CT 06459, USA  
\and
Department of Physics and Kavli Institute for Astrophysics and Space Research, MIT, Cambridge, MA 02139, USA
\and
National Astronomical Observatory of Japan, NINS, 2-21-1 Osawa, Mitaka, Tokyo 181-8588, Japan  
\and
Astronomical Institute, Czech Academy of Sciences, Fri\v{c}ova 298, 25165, Ond\v{r}ejov, Czech Republic  
\and
Astrobiology Center, NINS, 2-21-1 Osawa, Mitaka, Tokyo 181-8588, Japan  
\and
Center for Astronomy and Astrophysics, TU Berlin, Hardenbergstr. 36, 10623 Berlin, Germany  
\and
JST, PRESTO, 7-3-1 Hongo, Bunkyo-ku, Tokyo 113-0033, Japan 
\and
Department of Earth, Atmospheric and Planetary Sciences, MIT, 77 Massachusetts Avenue, Cambridge, MA 02139, USA 
\and
Institut de Ci\`encies de l'Espai (ICE, CSIC), Campus UAB,C/ de Can Magrans s/n, E-08193 Bellaterra, Spain  
\and
Institut d'Estudis Espacials de Catalunya (IEEC), C/ Gran Capit\`a 2-4, E-08034 Barcelona, Spain 
\and
Department of Theoretical Physics and Astrophysics, Masaryk University, Kotl\'{a}\v{r}sk\'{a} 2, 61137 Brno, Czech Republic 
\and
Astronomical Institute, Faculty of Mathematics and Physics, Charles University, Ke Karlovu 2027/3, 12116 Prague, Czech Republic  
}   

   \date{Received  20 March 2019; accepted xxx}

 
  \abstract
   {Although more than 2\,000 brown dwarfs  
   have been detected to date, mainly from direct imaging, their characterisation is difficult due 
   to their faintness and model dependent results. 
   In the case of   transiting brown dwarfs it is, however, possible to make direct high precision observations.
   }
   {Our aim is to investigate   the nature and formation of brown dwarfs   
   by  adding a new well-characterised   object, in terms of its mass, radius and bulk density,  
   to the currently  small    sample of less than 20 transiting brown dwarfs. } 
   {One brown dwarf candidate was found by the KESPRINT consortium
   when searching for exoplanets in the \emph{K2} space mission Campaign~16 field.
   We combined the \emph{K2} photometric data with a series of multi-colour photometric observations, imaging 
   and radial velocity measurements to rule out false positive scenarios and to determine the fundamental properties of the system.  
   }
   {We report the discovery and characterisation of a transiting brown dwarf  in a  5.17~day eccentric   orbit around 
    the slightly evolved   \stype~star \targetaa. We find a stellar  mass of \smasssouth~\Msun, a stellar radius of \sradiussouth~\Rsun, and an age of \age~Gyr. 
   The mass and radius of the companion brown dwarf are    \mpb~\mjup~and  \rpb~\rjup, respectively, 
   corresponding to a mean density of \denpb~\gc. 
  }
   {\targetaa b is a rare object that resides in  the  brown dwarf desert. 
   In the mass-density diagram for planets, brown dwarfs and stars, 
   we find that all giant planets and brown dwarfs follow the same trend from $\sim 0.3$~\mjup~to the turn-over 
   to hydrogen burning stars at $\sim73$~\mjup. \targetaa b falls   close to   the theoretical model for mature H/He dominated objects  
   in this diagram as determined by interior structure models, as well as   the empirical fit. 
 We argue that \targetaa b formed via gravitational disc instabilities in the outer part of the disc, followed by a quick migration. 
Orbital tidal circularisation may have started early in its history   for a brief period when the brown dwarf's radius was larger.  
The lack of spin--orbit synchronisation points to a weak stellar 
dissipation parameter ($Q^\prime_\star\gtrsim10^8$) which implies a circularisation timescale of $\gtrsim 23$~Gyr,  
or suggests an interaction between the magnetic and tidal forces of the star and the brown dwarf. 
}
   \keywords{Planetary systems -- Stars: fundamental parameters -- Stars:individual: \targetaa~ -- Techniques: photometric  -- Techniques: radial velocity 
               }

   \maketitle


\section{Introduction} \label{Section: introduction}
 The dividing 
line   between   gaseous giant planets (GPs) and brown dwarfs (BDs)     
is still unclear largely due
to the lack of well-characterised objects in this mass range.   
 BDs   have classically been regarded as objects in between large  planets and low-mass stars. Their
masses have been defined to be in the range $13-80$~\mjup~\citep{2001RvMP...73..719B}, sustaining  
deuterium burning through nuclear fusion for typically 0.1~million yrs, but    below the ignition limit of hydrogen at $75-80$~\mjup. 
Objects with masses above 65~\mjup~also  burn  lithium.  The   exact limits depend on models and 
internal chemical composition \citep{2014AJ....147...94D, 2011ApJ...727...57S, 2002A&A...382..563B}.
Another division between GPs and BDs  is 
based on formation: BDs are considered to  form like stars 
from gravitational instability on a dynamical timescale with the elemental abundance of the interstellar medium, while  GPs
form  on a longer timescale by core accretion with an enhanced metal abundance as compared
to their host star \citep{2014prpl.conf..619C}.  By this definition, the mass domains are overlapping since the minimum BD mass
is about 3~\mjup, and the maximum planet mass  can be as high as tens of \mjup. 
Others argue that BDs should not be distinguished from hydrogen-burning stars as 
they have more similarities to stars than planets  \citep{2018haex.bookE..95W}. 
 \citet{2015ApJ...810L..25H},  on the other hand,    suggested that BDs should be classified
as GPs instead of a separate class of its
own based on the mass-density relationship. They defined objects
within a mass range of $\sim 0.3 - 60$~\mjup~as the gaseous planet
sequence,  in analogy with the main sequence of stars. Objects below and above these limits were considered to be low-mass
planets and low-mass stars, respectively, although the upper limit
could be as high as 80~\mjup.  This was corroborated by \citet{2017ApJ...834...17C} who found that BDs follow the same
trend as GPs in the mass-radius diagram up to 80~\mjup. 

Although more than 2\,000 BDs have been detected
(e.g. \citet{2016A&A...589A..49S}, Johnston~2015\footnote{\url{http://www.johnstonsarchive.net/astro/browndwarflist.html}}) 
mainly by large-scale  direct imaging surveys, most 
of the detected BDs are free-floating, and only about 400 are found in bound systems 
at large distances from the primary star. Close BD companions to a main sequence star are very rare. 
Several surveys have showed that BDs in close orbits  ($<3$~AU) around main sequence FGKM stars 
have a much lower frequency than   GPs and close binaries 
\citep[e.g.][]{2000PASP..112..137M, 2006ApJ...640.1051G, 2011A&A...525A..95S}. 
This is commonly referred to as the BD desert   
 and may be a consequence of  different formation mechanisms for low- and high-mass BDs. 
BDs with masses \mbox{$35 \lesssim M \sin i \lesssim55$~\mjup}~and  orbital periods less than 100~days 
may represent the driest part of this desert \citep{2014MNRAS.439.2781M}. 
For objects in very close orbits,   $a < 0.2$~AU, \citet{2017A&A...608A.129T}  found  a paucity of 
 lower masses, $3-13$~\mjup. 
 
 It is evident that  many more well-characterised BDs are required to solve these issues. 
Characterisation from imaging is, however,  
 difficult since the objects are very faint, unless they are very young,  
and is heavily dependent on evolutionary models. 
In the case of eclipsing  
BDs the situation is different since   accurate determination of diameters is possible 
with photometric observations of the host star.   
Mass measurements are also relatively easy to perform   with high precision due
to the high masses of BDs. 
It is therefore possible to perform a model-independent characterisation of   individual 
BDs found by transit surveys combined with follow-up radial velocity (RV) measurements.

Space-based photometry allows excellent photometric precision and   long uninterrupted
observations \citep{2018haex.bookE..77F, 2018haex.bookE..79D,  2018haex.bookE..80B}.  
 This technique  has successfully been utilised to detect thousands of transiting exoplanets  by the 
 space missions \emph{CoRoT},       
\emph{Kepler} and its extension  \emph{K2}. 
The recently launched  \emph{TESS}  mission is expected to increase this number even further.  
The first discovery  of a transiting BD,  CoRoT-3b  \citep{2008A&A...491..889D},  was in fact made from space.  
The BD sample has since   grown   with additional detections from space, and also with  the ground-based surveys SuperWASP, 
 HATNet, MEarth,   
 and KELT.   
The sample of well-characterised objects   with  masses between   $\sim$10 and 80~\mjup~in 
bound systems is still, however, very small. Many more
are needed to investigate possible differences  between GPs and BDs.  
Using the classic 13~\mjup~limit between GPs and  BDs,   only 17  transiting BDs in bound systems around main sequence stars are known today.  
 A summary of  11 BDs,  five candidates, and two eclipsing BD binaries is found in Table~III.6.1 of \citet{2016cole.book..143C}. 
 Later discoveries of  six additional BDs have been made from space: Kepler-503\,b \citep{2018ApJ...861L...4C}, EPIC\,219388192\,b    \citep{2017AJ....153..131N}, 
  EPIC\,201702477\,b  \citep{2017AJ....153...15B},   
 and from the ground: WASP-128\,b 
\citep{2018MNRAS.481.5091H},   LP 261-75\,b \citep{2018AJ....156..140I}, and HATS-70\,b \citep{2019AJ....157...31Z}.

In this paper we report the independent discovery and observations of \targetaa b performed by the 
KESPRINT consortium. 
\citep[e.g.][]{2019MNRAS.484.3522H, 2019MNRAS.482.1807K, 2019MNRAS.484....8L, 2019A&A...623A..41P,   2018A&A...619L..10G}. 
\targetaa~was found in  the 
  \emph{K2} Campaign~16, and follow-up observations subsequently revealed that the object was
 the 18th transiting BD detected to date. 
We note that shortly before submitting this article, \citet{2019arXiv190303118C} publicly announced their 
discovery and RV observations of \targetaa b. 
We describe   the \emph{K2}  photometry  in 
Sect.~\ref{Section: K2 photometry and transit detection} and 
the   follow-up observations  in Sect.~\ref{Ground based follow-up}. 
We model the star in Sect.~\ref{Section: Stellar analysis},  
and  the transit and RVs  in Sect.~\ref{Section: modelling}.   
We end the paper with a discussion and conclusions in
Sect.~\ref{Section: Discussion} and \ref{Section: Conclusions}, respectively.


%
\begin{table}[!t]
\caption{Basic parameters for \targetaa $\tablefootmark{a}$.}
\begin{center}
\begin{tabular}{lll} 
\hline\hline
     \noalign{\smallskip}
Parameter    & Value   \\
\noalign{\smallskip}
\hline
\noalign{\smallskip}
\multicolumn{2}{l}{\emph{Main Identifiers}} \\
\noalign{\smallskip}
EPIC & 212036875    \\
2MASS &  J08584567+2052088 \\
WISE &  J085845.66+205208.4 \\
TYC & 1400-1873-1 \\
UCAC & 555-045746 \\
GAIA DR2 &	684893489523382144	\\
\noalign{\smallskip}
\hline
\noalign{\smallskip}
\multicolumn{2}{l}{\emph{Equatorial coordinates}} \\
\noalign{\smallskip} 
$\alpha$(J2000.0) & $08^{\rm h}\,58^{\rm m}\,45\overset{\rm s}{.}67$     \\
$\delta$(J2000.0) & +20$^{\circ}\,52\arcmin\,08\farcs78$    \\
\noalign{\smallskip}
\hline
\noalign{\smallskip}
\multicolumn{2}{l}{\emph{Magnitudes}} \\
$B$ (Johnson) &   $11.654\pm0.113$     \\
$V$ (Johnson)  & $10.950\pm0.095$      \\
$G$ (Gaia)	&   $10.9148\pm0.0009$ \\
$Kepler$ & 10.937      \\
 $g$ & $12.257\pm0.050$ \\
 $r$ & $10.918\pm0.060$ \\
  $i$ & $10.800\pm0.070$ \\
$J$  &   $10.042\pm0.022$     \\
$H$    &   $9.843\pm0.024$      \\
$K$   &     $9.774\pm0.018$   \\
\noalign{\smallskip}
\hline
\noalign{\smallskip}  
Parallax   (mas) &\parallaxgaia\,   \\  
Systemic velocity (\kms) & \velgaia   \\   
$\mu_{RA}$ (mas~yr$^{-1}$) & \pmra   \\   
$\mu_{Dec}$ (mas~yr$^{-1}$) & \pmdec   \\  
    \noalign{\smallskip} \noalign{\smallskip}
\hline 
\end{tabular}
\tablefoot{
\tablefoottext{a}{From the Ecliptic Plane Input Catalogue \citep[EPIC; ][]{2016ApJS..224....2H} \url{http://archive. stsci.edu/k2/epic/search.php} and
the Gaia~DR2 archive \url{http://gea.esac.esa.int/archive/}.} 
}
\end{center}
\label{Table: Star basic parameters}
\end{table}

  \begin{figure*}[!ht]
 \centering
  \resizebox{\hsize}{!}
            {\includegraphics{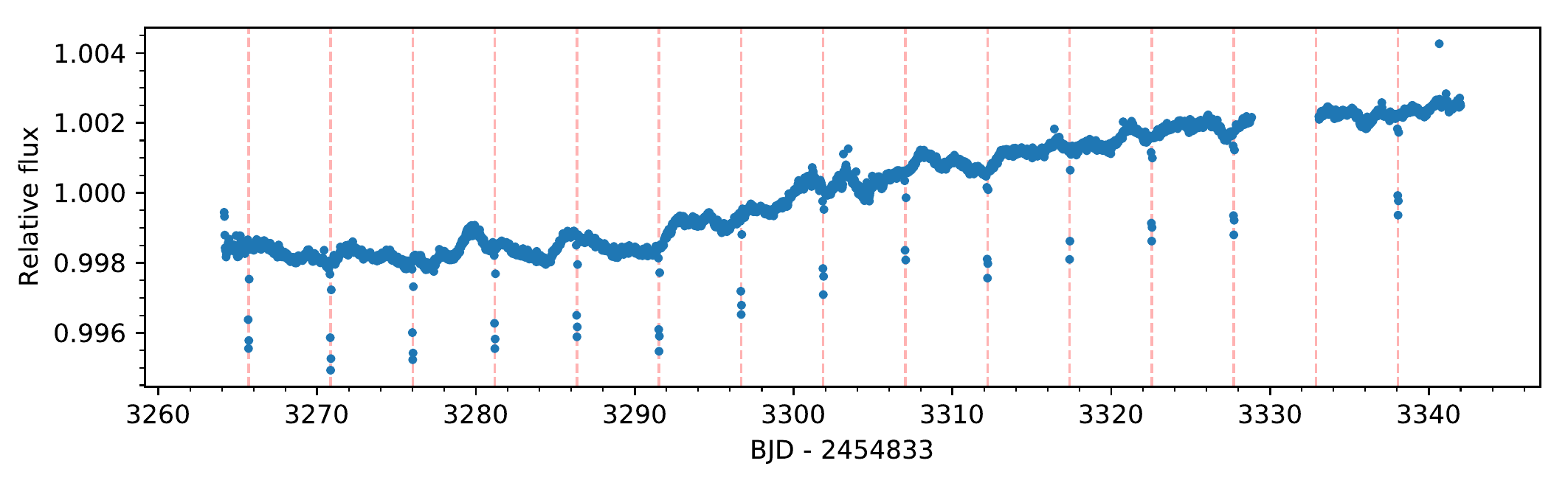}}
   \caption{The pre-processed Vanderburg \emph{K2} light curve of \targetaa. The dashed vertical lines mark the 
   14 narrow and shallow   brown dwarf transits used in the analysis. 
   We also mark a missing transit located in a gap of the light curve. 
   The  broader periodic variation of approximately 0.07~\% is caused by stellar activity.}
      \label{Figure: full lightcurve}
 \end{figure*}
%

 \section{\emph{K2} photometry and transit detection} \label{Section: K2 photometry and transit detection}
Between 7~Dec~2017 and 25~Feb~2018,  
the \emph{Kepler} space telescope   monitored 35\,643 objects 
in the long (29.4~min) cadence mode, and  
131 objects with short (1~min) cadence in the direction 
towards ($J2000$) 
\mbox{$\alpha = 08^h54^m 50\farcs3$} and \mbox{$\delta = +01^\circ14\arcmin 06\farcs 0$}  
(the \emph{K2} Campaign~16\footnote{\url{https://keplerscience.arc.nasa.gov/k2-data-release-notes.html\#k2-campaign-16}}).  
  The  data  of Campaign~16 was downloaded    from   the Mikulski Archive for 
Space Telescopes\footnote{\url{https://archive.stsci.edu/prepds/k2sff/}} (MAST). 
We followed the procedure described in \citet{2019MNRAS.482.1807K} and  searched 
for periodic signals  in the photometric data using the {\tt{EXOTRANS}} software \citep{Grziwa2012}. 
The software utilises 
wavelet-based filters to remove stellar variability and instrument systematics, and  
 a modified {\tt{BLS}} 
\citep[Box-fitting Least Squares; ][]{Kovacs2002} 
algorithm, improved by implementing 
optimal frequency sampling \citep{2014A&A...561A.138O}, to detect the most significant transits. 
Periodic signals were detected 
in the light curve of the  \stype~star \targetaa~with an orbital period of  \mbox{5.17~days},   
a mid-transit time $T_0 = 3265.68$~days~\mbox{(BJD - 2454833)}, and a depth of $\sim0.4~\%$. 
The   pre-processed   
Vanderburg\footnote{\url{https://www.cfa.harvard.edu/~avanderb/k2c16/ep212036875.html}}   
light curve is  shown  in Fig.~\ref{Figure: full lightcurve}. 
The depth  is consistent with a   Jupiter-sized planet, 
although the nature of the planet candidate had to await radial velocity follow-up. 
We found no signs of even-odd   depth variations or a secondary eclipse  within $1~\sigma$, 
which  is a first step to excluding binaries. We thus proceeded with a follow-up campaign to characterise the \targetaa~system. 

The basic parameters of the star are listed in Table~\ref{Table: Star basic parameters}.

 \begin{figure*}[!t]
 \centering
  \resizebox{\hsize}{!}{
   \includegraphics[width=\linewidth]{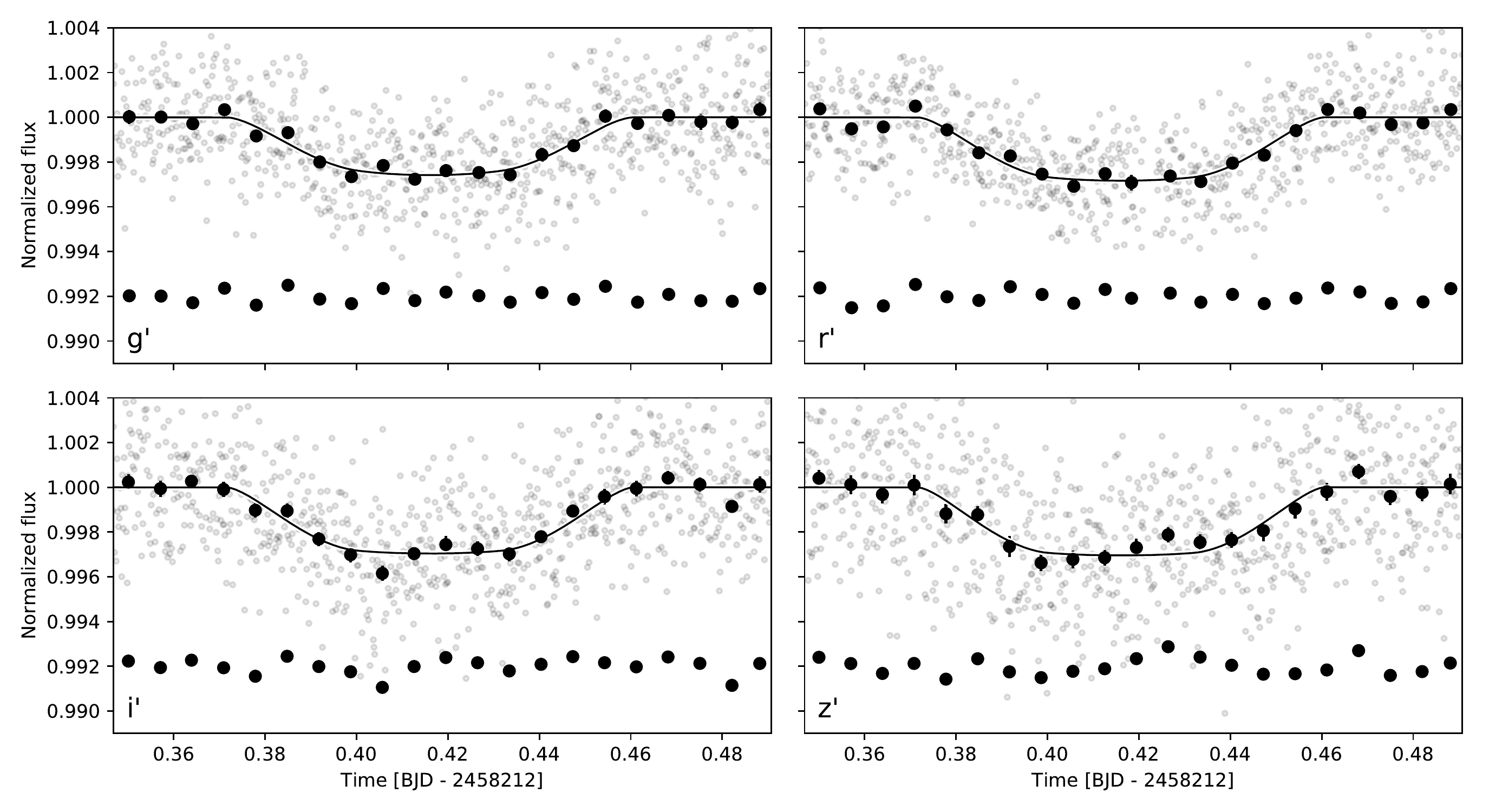}}
   \caption{Transit light curves of \targetaa b obtained with MuSCAT2 (TCS),  
   in $g'$ (upper-left), $r'$ (upper-right), $i'$ (lower-left), and $z'$
(lower-right) filters, respectively. The black solid line is the
best-fit model accounting for the de-trending and transit components.
The residuals are plotted in the lower portion of each respective plot. 
}
      \label{Figure: Muscat 2}
 \end{figure*}
%


\section{Ground-based follow-up} \label{Ground based follow-up}

 We performed a series of follow-up observations with 
\emph{(i)} 
multi-colour photometric observations   to rule 
out eclipsing binary false-positives (Sect.~\ref{Section: muscat2}); \emph{(ii)}   reconnaissance spectra observations 
to remove candidates with rapidly rotating 
stars, double-lined binaries and blends of spectral components (Sect.~\ref{Subsection: KEA measurements}); \emph{(iii)} 
RV follow-up  
to obtain the BD mass and co-added spectra needed 
for stellar spectral modelling (Sect.~\ref{Subsection: RV measurements}); 
and \emph{(iv)} high   resolution 
adaptive optics (AO; Sect.~\ref{Subsubsection: AO imaging}) and  
speckle imaging (Sect.~\ref{Subsubsection: NESSI imaging})    
to search for contaminant 
stars that may be background or foreground stars, or physically bounded 
eclipsing binaries whose light may be diluted by the target star and 
generate transit-like signals. 
Speckle and AO observations are fundamentally different techniques;  
NESSI speckle probes the inner region ($<0.2\arcsec$) around the target star at optical wavelengths, 
while AO, achieves a much higher contrast  in the  $0.2\arcsec - 1.0\arcsec$ region  in the near infrared. 
 These regions are not possible to explore with the \emph{K2} data with a sky-projected pixel size of  4\arcsec.

\subsection{MUSCAT2}\label{Section: muscat2}

We observed a full transit of  \targetaa b  with MuSCAT2 at the Carlos
Sanchez Telescope (TCS) on the night of 3~April 2018. MuSCAT2 is a
4-colour imager that allows for simultaneous observations in $g'$, $r'$, $i'$, and
$z'$  \citep{2018arXiv180701908N}. The observations started at 20:15~UT
and ended at 23:30~UT, covering the full transit and some pre- and post 
transit baselines. The night was clear, with variable seeing between 1\arcsec
and 2\arcsec. Exposure times were set to 5~s in all channels.

The differential photometry and transit light curve analysis were carried out
with a dedicated MuSCAT2 pipeline. The photometry follows standard aperture photometry
practices: we calculated an astrometric solution for each frame
using an offline version of astrometry.net \citep{Lang2010},
and retrieved the photometry for a set of comparison stars and aperture sizes.
 
The transit modelling continued by first choosing a set of optimal apertures that minimise the
relative light curve point-to-point scatter. Next, we jointly fitted a transit model with a linear
baseline model (a linear model in sky level, airmass, seeing, and CCD position variations) to
the four light curves using PyTransit and LDTk \citep{Parviainen2015, Parviainen2015b}.
Finally, we swapped the linear baseline model to a Gaussian process-based model  with the final
kernel consisting of a \mbox{product} of squared exponential kernels for all the covariates, 
and carried out MCMC sampling to obtain an estimate of the model parameter posterior distribution.
The final light curves are shown in Fig.~\ref{Figure: Muscat 2}.
The transit model allows for colour-dependent variations in transit depth due to blending by
an unresolved source, and our analysis allows us to rule out any significant contamination
that would affect the parameter estimates derived from the transit photometry.

\subsection{Reconnaissance spectra with Tull} \label{Subsection: KEA measurements}
 
On 5 April 2018 we obtained  a  reconnaissance spectrum of \targetaa~with the Tull spectrograph
at the 2.7~m telescope at McDonald Observatory. The high-resolution \mbox{($R \approx 60\,000$)} spectrum   
was reduced using standard {\tt{iraf}}
routines. 
We derived a first estimate of the stellar spectroscopic parameters using the code {\tt{Kea}} \citep{Endl2016}: 
\mbox{\teff = $6380 \pm 58$~K}, \mbox{\feh = $-0.21 \pm 0.03$~dex},
 \mbox{$\log(g_\star) =  4.25 \pm 0.14$}~(cgs), and 
\mbox{\vsini  = $11.9 \pm 0.3$~\kms}. 
We found no evidence of   a double-lined binary or any blends of spectral components.

\subsection{Radial velocity follow-up with FIES} \label{Subsection: RV measurements}
The RV follow-up was performed with   FIES  
\citep[the FIbre-fed \'Echelle Spectrograph;][]{Telting2014, Frandsen1999} 
mounted on the 2.56~m Nordic Optical Telescope 
 (NOT) at the Roque de los Muchachos Observatory. 
 We observed nine high-resolution ($R \approx 67\,000$)  
 spectra between 9~April~and~22~May~2018 as 
 part of our CAT and TAC programmes 57-015, 57-206, and 57-210, and OPTICON program 
 2018A-044. To account for the RV offset caused by a major instrument refurbishment 
 that occurred on 30~April 2018, we treated the RV taken between 9 and 26~April, and between 
 6 and 8~May as two independent data-sets. In addition, 14 intermediate-resolution ($R \approx 47\,000$) 
  FIES spectra were also acquired between 12~May~2018 and 
 26~Feb~2019, as part of the OPTICON programme 2018B-052 and the Spanish-Nordic 
 programme 58-301. Depending on the sky conditions and scheduling constraints, 
  we  set the exposure times to 1800 -- 3600~s for both resolutions. 
 To trace the RV drift of the instrument we followed 
the strategy outlined in \citet{Gandolfi2015} and \citet{2010ApJ...720.1118B}   and 
 bracketed the science exposures with long-exposed (60--90~s) ThAr spectra. 
 We used the standard IRAF and IDL routines to reduce the data.   
  The S/N ratio of the extracted spectra ranges between $\sim$35 and 75 per pixel at 
 5500~\AA. Radial velocities were extracted via multi-order cross-correlations with the 
 spectrum of the RV standard star HD\,168009, for which we adopted an absolute RV 
 of -64.650~\kms\citep{1999ASPC..185..367U}.

The FIES RVs  are listed in Table~\ref{Table: RV measurements}. 
Figure~\ref{Fig: RV periodogram}  shows the generalised Lomb-Scargle periodogram of the offset-corrected Doppler 
measurements (combined by subtracting the systemic velocities   listed in 
Table~\ref{Table: Orbital and planetary parameters}). We found a very significant peak  
at the orbital frequency of the transiting brown dwarf with a false alarm 
probability FAP\, $\ll 10^{-6}$, proving that the Doppler reflex motion of the star 
induced by the orbiting companion is clearly detected in our data.

\label{Fig: RV periodogram}

\subsection{Subaru/IRCS AO imaging} \label{Subsubsection: AO imaging}
 
 In order to obtain high-contrast, high-resolution images of \targetaa, 
we performed AO imaging with the InfraRed Camera and Spectrograph  
\citep[IRCS,][]{2000SPIE.4008.1056K} atop the Subaru 8.2~m telescope on 14~June 2018. 
The target star was used as a natural guide and  AO correction was applied to obtain high-contrast
$K^\prime$-band images of the target. 
We used the fine-sampling mode ($1~\mathrm{pix} \approx 21~\mathrm{mas}$), 
and implemented a five-point dithering to minimise the impacts of bad pixels and cosmic rays.

We reduced the raw frames with a standard procedure described in \citep{2016ApJ...820...41H} 
to produce an aligned and combined image of \targetaa.
The full width at the half maximum of the
co-added target image was $0\farcs089$  suggesting that the AO correction worked well
for this target. 
As shown in the inset of Fig.~3, EPIC 212036875 exhibits no nearby source. 
Following   \citet{2018AJ....155..127H} we estimated 
the detection limit of possible nearby sources by computing  a $5~\sigma$ contrast curve    drawn in Fig.~3.
The  achieved contrast is $\Delta m_{K^\prime}>7$ mag beyond $0\farcs5$ from \targetaa. 


%
  \begin{figure}[!ht]
            {\includegraphics[scale=1.18]{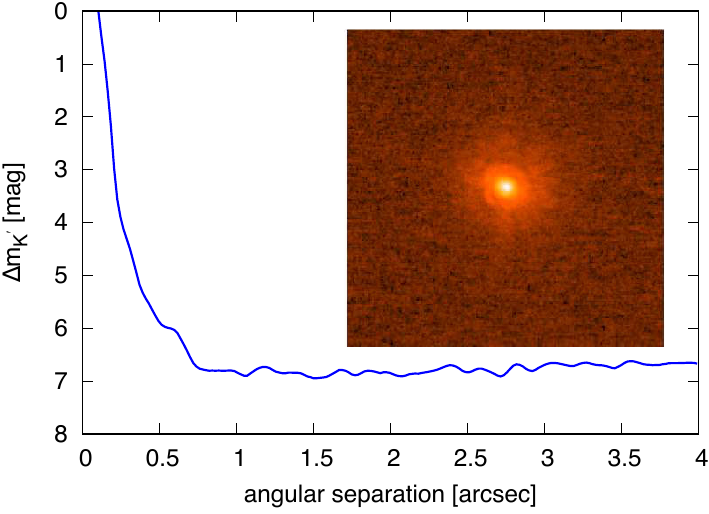}}
   \caption{IRCS/Subaru AO-imaging in the $K^\prime$-band  and 5~$\sigma$ magnitude contrast curve as a function of angular
   separation from \targetaa. The inset shows the $16\arcsec \times 16\arcsec$  
       saturated image. Northeast is up and to the left.}
      \label{Figure: AO image}
 \end{figure}
  \begin{figure}[!ht]
            {\includegraphics[scale=0.82]{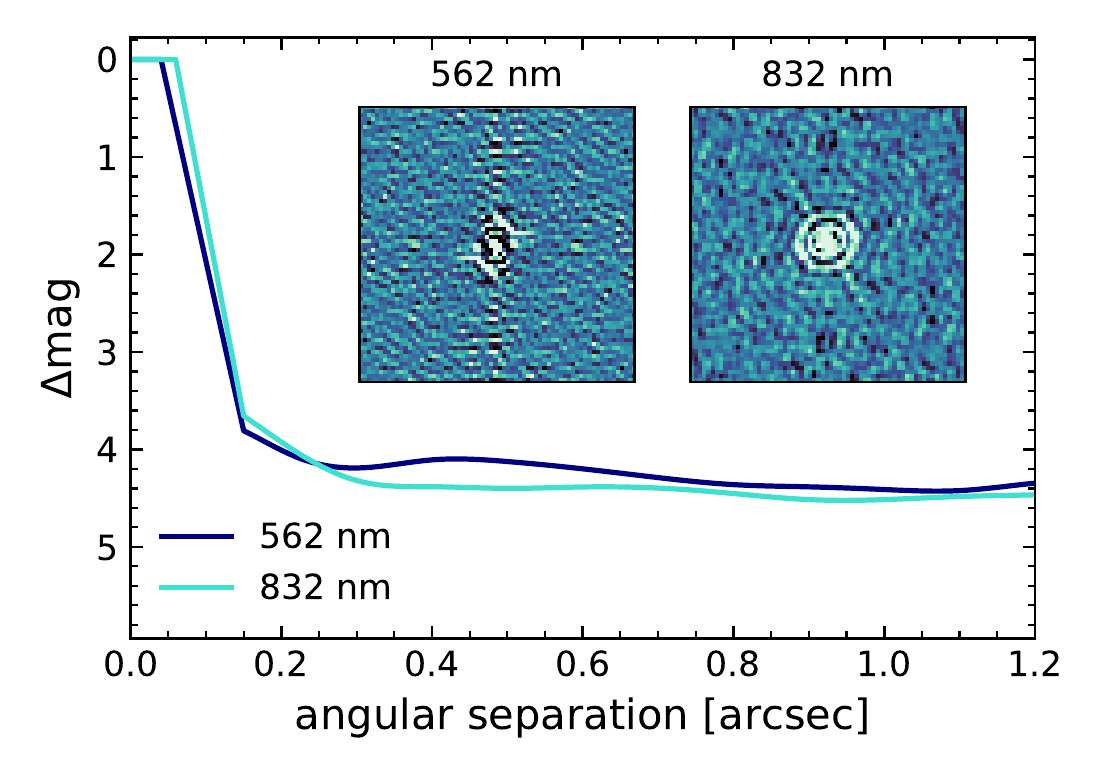}}
 \caption{NESSI/WIYN   speckle interferometry reconstructed images in the $r$-  and  $z$-narrowbands  and
  $5~\sigma$   contrast curves. The inset images are $1\farcs2\times1\farcs2$. Northeast is up and to the left.}
      \label{Figure: speckle imaging}
 \end{figure}
%


 \subsection{NESSI imaging} \label{Subsubsection: NESSI imaging}
 
 To further constrain the presence of stellar companions at close separations, 
we conducted speckle imaging of \targetaa~using the 
NASA Exoplanet Star and Speckle Imager  \citep[NESSI;][]{2018SPIE10701E..0GS} at the WIYN 3.5~m telescope  on 19~June~2018 (program ID 2018A-0181).
NESSI operates simultaneously in two bands centred at 562~nm \mbox{($r$-narrowband)} and 832~nm 
\mbox{($z$-narrowband)}. 
We collected and reduced the data 
following the procedures described by \citet{2011AJ....142...19H}, 
yielding $4\farcs6\times4\farcs6$ reconstructed 
images of the host star (the inset shows the central  $1\farcs2\times1\farcs2$  
in Fig.~\ref{Figure: speckle imaging}). 
We did not detect any secondary sources in the reconstructed images. The 
$5~\sigma$ detection limits are shown in Fig.~\ref{Figure: speckle imaging}; the contrast is \mbox{approximately} 4.5~mag
beyond  $0\farcs3$ from \targetaa~in both images. 

We  used the NESSI and Subaru magnitude limits to estimate   limits on companion masses vs. separation 
and find that massive companions are excluded outside $\sim$100~AU (Fig.~\ref{Figure: contrast curve masses}). 
 

\section{Stellar analysis} \label{Section: Stellar analysis}

 \subsection{Spectral analysis} \label{Subsection: spectral analysis}

Before we modelled  the  BD, we first   computed the absolute 
mass and radius of the host star.  
In order to obtain the   stellar parameters needed in the stellar models,     
we used the spectral analysis package {\tt{SME}} \citep[Spectroscopy Made Easy;][]{vp96, pv2017}.
This software calculates synthetic stellar spectra from grids of atmosphere models which are then  fitted to the observations
using a $\chi^2$-minimising procedure.  Here we specifically used the  {\tt{ATLAS12}} model spectra \citep{Kurucz2013}, and 
{\tt{SME}} version 5.22 to model our co-added FIES spectra.
We followed 
well established methods  described in  \citet{2017A&A...604A..16F} and \citet{2018A&A...618A..33P}  
to compute \teff, \logg, \vsini, and abundances. The micro- and macro-turbulent velocities, \vmic~and \vmac, 
were fixed using the   calibration equations for Sun-like stars from \citet{Bruntt2010b} and \citet{Doyle2014}, 
respectively.  
The line lists were taken  from  the 
Vienna Atomic Line Database\footnote{{\url{http://vald.astro.uu.se}}} \citep{Ryabchikova2015}.

Our results obtained with SME \mbox{(\teff \, = \, \steffsme~K)} are in   agreement with the values
listed in  the Gaia DR2 archive \mbox{(\teff \, = \, \steffgaia~K)} and 
EPIC \mbox{(\teff \, = \, \steffepic~K)}, and  are also consistent with the   
Kea results from the Tull reconnaissance spectra in 
Sect.~\ref{Subsection: KEA measurements} \mbox{(\teff \, = \, \steffkea~K)}.  The resulting \teff~and the  luminosity  in the 
Gaia DR2 archive  implies a spectral type of \stype. 
All final   results are listed in Table~\ref{Table: Final stellar parameters}.

\subsection{Stellar mass and radius} \label{Subsection: stellar mass and radius}
We used the     \citet{2011MNRAS.417.2166S} calibration equations to compute 
the stellar mass and radius. 
These empirical relations, based on data from eclipsing binaries, are valid for masses up to 3~\Msun~and
account for metal abundance and evolution.
It provides the advantage of using the stellar density  which   has a higher precision than   \logg~since it  
is derived from the  the transit light curve.  
Additional input parameters are \teff~and \feh.

We also  compared the Southworth results   
with    several other, independent methods. The first is the \citet{Torres2010} calibration equations based on a different set 
 of eclipsing binaries, as well as interferometrically determined stellar diameters. 
 The input parameters are \teff, \logg, and \feh. 
We   further applied the Bayesian 
{\tt{PARAM\,1.3}}\footnote{\url{http://stev.oapd.inaf.it/cgi-bin/param_1.3}} model tool tracks \citep{daSilva2006} 
with the  {\tt{PARSEC}} isochrones   \citep{2012MNRAS.427..127B}  and the
 apparent visual magnitude,  \teff, \feh, and the  parallax as input.   
 The derived age and \logg~from  {\tt{PARAM\,1.3}} are  
\mbox{\age~Gyr} and \mbox{\sloggp~(cgs)}, respectively. 
 Finally, when we compared the derived mass and radius to a typical  \stype~star, we noted that
\targetaa~seems to be   slightly evolved,  in line 
with a typical life time of about \mbox{$\sim 7$~Gyr}. 
\targetaa b is one of only two BDs where the age can
be determined relatively precisely, due to its evolutionary state. The other BD with a well
determined age is  EPIC~219388192b  
\citep{2017AJ....153..131N} which is a member of Ruprecht~147, 
the oldest nearby open cluster association.

All models  are in excellent agreement with each other.  
The results from all models are listed in Table~\ref{Table: Comparison stellar parameters}, and the final 
adopted stellar parameters are listed in Table~\ref{Table: Final stellar parameters}.

\begin{table}
 \centering
 \caption{Adopted  stellar parameters of \targetaa.}
\begin{tabular}{ll}
 \hline\hline
     \noalign{\smallskip}
 Parameter &\targetaa~    \\
    \noalign{\smallskip}
     \hline
\noalign{\smallskip} 
Effective temperature \teff$\tablefootmark{a}$~(K)\dotfill &\steffsme  \\ \noalign{\smallskip}
Surface gravity $\log(g_\star)\tablefootmark{a, b}$   (cgs)        \dotfill & \sloggMgsme  \\  \noalign{\smallskip}              

Metallicity [Fe/H]$\tablefootmark{a}$ (dex)   \dotfill &  \sfehsme  \\ \noalign{\smallskip}
Metallicity [Ca/H]$\tablefootmark{a}$ (dex)   \dotfill &  \scahsme  \\ \noalign{\smallskip}
Metallicity [Na/H]$\tablefootmark{a}$ (dex)   \dotfill &  \snasme  \\ \noalign{\smallskip}
Metallicity [Mg/H]$\tablefootmark{a}$ (dex)   \dotfill &  \smghsme  \\ \noalign{\smallskip}

Rotation velocity \vsini$\tablefootmark{\,a,c}$~(\kms)  \dotfill& \svsini      \\  \noalign{\smallskip}
Microturbulent $V\tablefootmark{d}$~(\kms)  & \svmic \\  \noalign{\smallskip}
Macroturbulent $V\tablefootmark{e}$~(\kms)  & \svmac \\  \noalign{\smallskip}

Mass $M_\star\tablefootmark{f}$  (\Msun)  \dotfill & \smasssouth  \\ \noalign{\smallskip}
Radius $R_\star\tablefootmark{f}$   (\Rsun)  \dotfill &  \sradiussouth \\ \noalign{\smallskip}
Density  $\rho_\star\tablefootmark{g}$   (g~cm$^{-3}$)\dotfill & \denspyaneti  \\ \noalign{\smallskip}

Luminosity $L_\star\tablefootmark{h}$    (\Lsun)\dotfill & \lstargaiaarchive    \\ \noalign{\smallskip}
 
Spectral type   \dotfill&  \stype  \\ \noalign{\smallskip}

Rotation period$\tablefootmark{i}$ (days) \dotfill&  \prot   \\ 

Age$\tablefootmark{j}$  (Gyr) \dotfill&  \age \\ 
    \noalign{\smallskip} \noalign{\smallskip}
\hline 
\end{tabular}
\tablefoot{
\tablefoottext{a}{From  {\tt{SME}} modelling.}
\tablefoottext{b}{Modelled using \ion{Mg}{I}.  The \ion{Ca}{I} model gives $\log(g_\star)$ = \sloggCasme~(cgs).}
\tablefoottext{c}{The projected stellar rotation speed of its surface.}
\tablefoottext{d}{Fixed with the empirical calibration  by  \citet{Bruntt2010b}.}
\tablefoottext{e}{Fixed with the empirical calibration  by \citet{Doyle2014}.}
\tablefoottext{f}{\citet{2011MNRAS.417.2166S} calibration equation.} 
\tablefoottext{g}{Density from pyaneti transit modelling in Sect.~\ref{Section: modelling}. Density   from adopted stellar mass and
radius is \mbox{\densMR~g~cm$^{-3}$}.} 
\tablefoottext{h}{Gaia DR2 archive.}
\tablefoottext{i}{From the generalised Lomb-Scargle periodogram.} 
\tablefoottext{j}{ {\tt {PARAM 1.3}}.} 
}
 \label{Table: Final stellar parameters}
\end{table}

\begin{table}
 \centering
 \caption{Stellar mass and radius of \targetaa~as derived from different methods.
 The  typical values   for a  \stype~star are listed as comparison. 
 }
\begin{tabular}{lrr}
 \hline\hline
     \noalign{\smallskip}
Method  & \mstar &        \rstar    \\
 &  (\Msun)  & (\Rsun)     \\
    \noalign{\smallskip}
     \hline
\noalign{\smallskip} 

Southworth$\tablefootmark{a}$    
& \smasssouth& \sradiussouth    \\

 Torres$\tablefootmark{b}$    
& \smasst&\sradiust    \\
   
  {\tt {PARAM 1.3}}      
 & \smassp    & \sradiusp  \\

 Gaia DR2$\tablefootmark{c}$ 
& \ldots    &  \srgaia  \\   

EPIC$\tablefootmark{d}$ 
& \smepic    &  \srepic  \\  

   Spectral type$\tablefootmark{e}$  \stype       
 & \smassspectral& \sradiusspectral    \\ 

      \noalign{\smallskip} \noalign{\smallskip}
\hline 
\end{tabular}
\tablefoot{
\tablefoottext{a}{\citet{2011MNRAS.417.2166S} calibration equations.}
\tablefoottext{b}{\citet{Torres2010} calibration equations.}
\tablefoottext{c}{Gaia DR2 archive.}
\tablefoottext{d}{The K2 Ecliptic Plane Input Catalog.}
\tablefoottext{e}{\citet{2000asqu.book.....C}.}
}
 \label{Table: Comparison stellar parameters}
\end{table}


\subsection{Stellar  rotation period} \label{Section: stellar activity}

The \emph{K2} light curve of \targetaa\ displays periodic and quasi-periodic photometric variations 
with a semi-amplitude of $\sim$0.07\%. These are superimposed on a long-term photometric trend 
with a peak-to-peak amplitude of $\sim$0.4\% (Fig.~\ref{Figure: full lightcurve}), which we attributed 
to the slow drift often present in \emph{K2} data \citep{VandJohn2014}. 
Given the spectral type of  the host star,  
the periodic and quasi-periodic variability is   likely induced by magnetically 
active regions carried around by stellar rotation.

We used the generalised Lomb-Scargle (GLS) periodogram \citep{2009A&A...496..577Z} and the 
auto-correlation function (ACF) method \citep{2014ApJS..211...24M} to estimate the rotation period of 
the star. Prior to computing the GLS periodogram and the ACF, we masked out the transits and 
removed the long-term trend by dividing the out-of-transit light curve by the best-fitting 4th-order 
cubic spline (Fig.~\ref{Fig: light curve periodogram}, upper panel). The GLS periodogram of the corrected light curve 
(Fig.~\ref{Fig: light curve periodogram}, middle panel) 
shows a very significant peak at $f=0.14$~d$^{-1}$ ($P_\mathrm{rot} =  7.2$~days) with  
FAP\,$\ll 10^{-6}$, estimated from the bootstrap method    \citep{1997A&A...320..831K}. 
The ACF of the light curve (Fig.~\ref{Fig: light curve periodogram}, lower panel) shows correlation peaks at 
$\sim$7, 14, 21, 28 days. We interpreted the peak at $\sim$7\,days as the rotation period of the star 
and the peaks at $\sim$14, 21, and 28 days as its first, second, and third harmonics, respectively. 
By fitting a Gaussian function to the highest peak of the GLS periodogram, we derived a rotation 
period of $P_\mathrm{rot}=7.2\pm0.5$\,days. Assuming that the star is seen almost equator-on 
\mbox{($\sin i_\star \approx 1$)}, the spectroscopically derived rotational velocity \vsini\ and the 
stellar radius imply a rotation period of $6.6 \pm 0.9$~days, in very good agreement with our results. 
The orbital period of the brown dwarf    is thus within 7~\%   
to a 3:2 commensurability with the stellar rotation period.
 
  We used 
the formula from \citet{2007AJ....133.1828W}  to constrain   $i_\star$   
(the inclination of the stellar spin axis relative to the sky plane), 
and found 
\mbox{$\sin i_\star = V \sin i_\star P_\mathrm{rot}/(2 \pi  R_\star) \approx 1.09 \pm 0.17$}. 
The value with $\sin i_\star > 1$ was rejected as unphysical and we determined  a lower 
bound of $i_\star$   to \istar$^{\circ}$ with 1~$\sigma$ confidence.

Since the \vsini~of \targetaa~is relatively high, the Rossiter-McLaughlin (RM) 
effect  could  be   measured with current state-of-the-art spectrographs, 
mounted on \mbox{8--10~m} class telescopes, using either RV RM or Doppler tomographic methodology.
A first order estimate of the amplitude of the RM effect is $\sim$16~m~s$^{-1}$
using the equation \mbox{$\Delta V = (R_\mathrm{bd}/R_\star)^2 \times \sqrt{1-b^2} \times V \sin i_\star$} 
\citep{2010exop.book...55W, 2018haex.bookE...2T}. 
Note, however, that with a large impact parameter the actual amplitude of the RM 
effect is a strong function of the angle between the sky projections of 
the stellar spin axis and the orbit normal ($\lambda$), implying that the actual RM 
amplitude could vary substantially from the above estimate.

 Apart from the independent measurements of \vsini~and $\lambda$,  
the measurement of the RM effect, together with the \Prot~and \vsini~measurements to constrain the inclination of the 
stellar rotation axis, would also allow a constraint upon the 
misalignment angle,  $\psi$ (the 3-D obliquity angle between the stellar spin 
axis and the orbital axis). 
Measuring the spin-orbit misalignment of \targetaa b would be valuable because there are only 
a handful of such measurements available for transiting BDs \citep{Triaud09, 2012ApJ...761..123S, Triaud13, 2019AJ....157...31Z}. 
Furthermore, this object 
is the only one of these for which the full 3-D spin-orbit angle is measurable, allowing better constraints 
on the system architecture. Finally,  all of the other objects observed to date have circular orbits, 
unlike \targetaa b; measuring the spin-orbit misalignment will enable a full 
dynamical characterisation of this system, which will have consequences for our understanding of 
how the system formed (see Sect.~\ref{subsection: Tidal evolution}).


\section{Transit and Radial Velocity modelling} \label{Section: modelling}

We used the well tested and publicly available  PYTHON/FORTRAN {\tt{pyaneti}}\footnote{\url{https://github.com/oscaribv/pyaneti}} 
\citep{2019MNRAS.482.1017B} package   to carry out simultaneous modelling of both the
 \emph{K2} light curve and the  FIES RV measurements. 
 The code uses Markov chain Monte Carlo (MCMC) methods based on Bayesian analysis
 and has successfully been used by us in e.g. \citet{2019ApJ...876L..24G} and \citet{2018MNRAS.475.1765B}.  
 In preparation for the modelling, the light curve was detrended with the  {\tt{exotrending}} \citep{2017ascl.soft06001B} 
 code. This procedure reduces the flux variations   of any long-term
systematic or instrumental trends. 
 Each of the 14 transits was cut out of the light curve, and four hours around 
each   transit  were masked to ensure that no in-transit data was used in the process, before fitting a second 
order polynomial to the remaining out-of-transit data.

 
%
 \begin{figure}[!t]
 \centering
  \resizebox{\hsize}{!}
   {\includegraphics{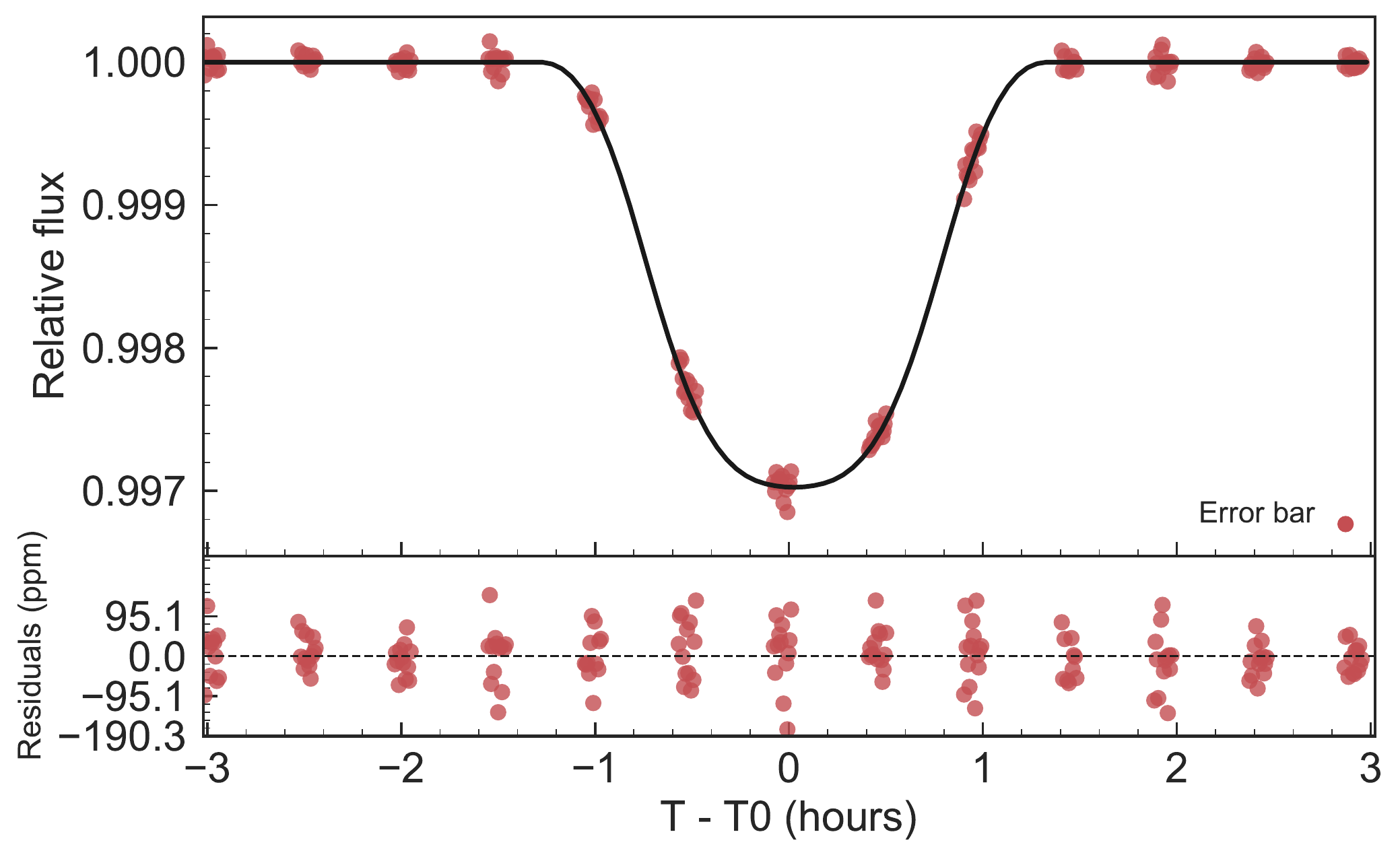}}
   \caption{Transit light curve folded to the orbital period of \targetaa b. The  
   \emph{K2} photometric data is indicated with the red points, and the 
   best-fitted transit model  with the solid black line. 
    The residuals of the fit are shown in the lower panel.}
   \label{Figure: pyaneti transit model}
 \end{figure}
 \begin{figure}[!hbt]
 \centering
  \resizebox{\hsize}{!}{
   \includegraphics[width=\linewidth]{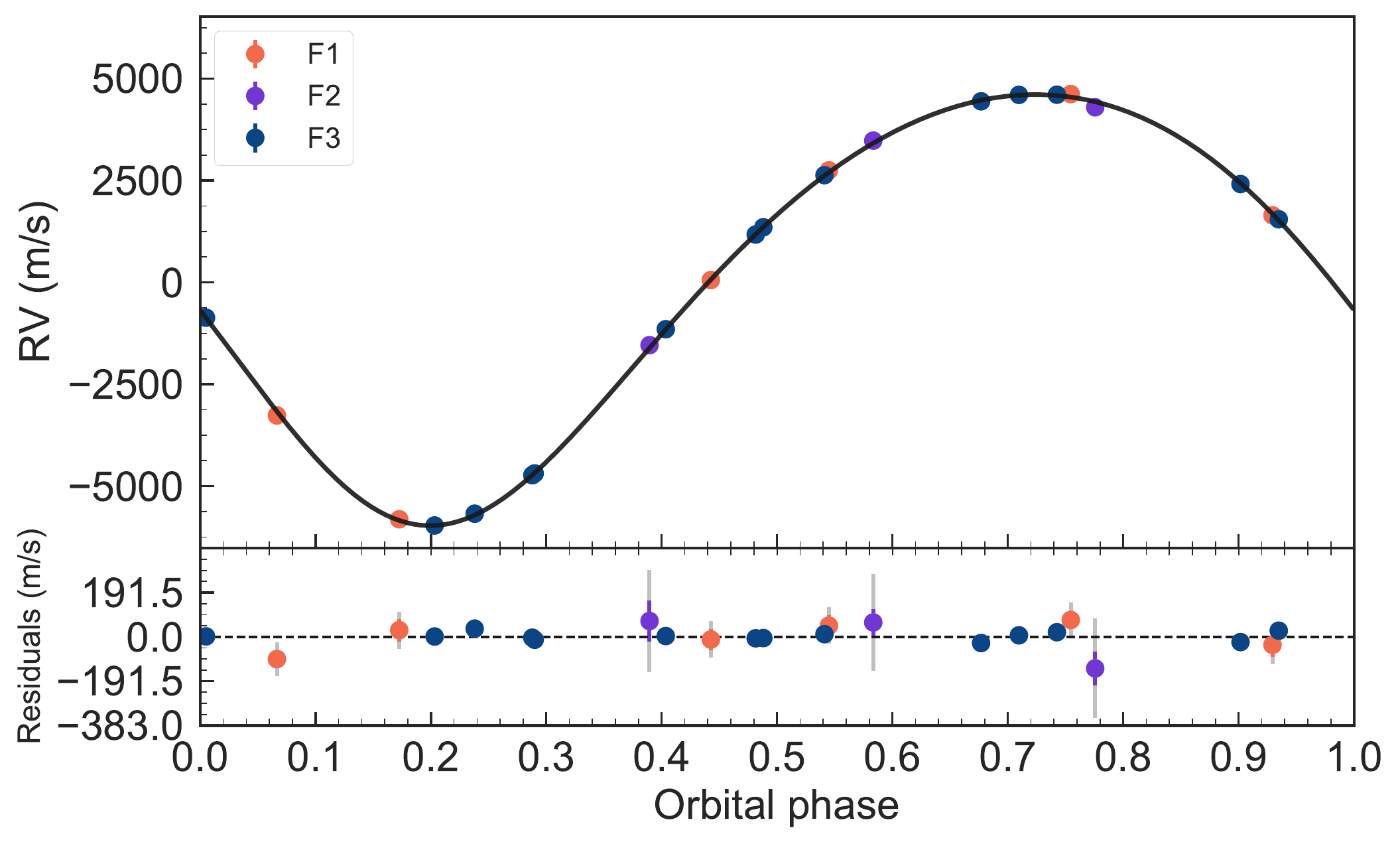}}  
   \caption{Radial velocity curve  of \targetaa~phase folded to the orbital period of the brown dwarf. 
  The different colours mark the different FIES setups, and the 
  best-fitted RV model is indicated with the  solid black line. 
   The residuals of the fit are shown in the lower panel. The coloured error bars are without jitter, and the  grey error
   bars  includes the jitter.}
      \label{Figure: pyaneti rv model}
 \end{figure}
%


Following \citet{2018A&A...612A..95B}, we  fitted a Keplerian orbit to the RV data with 
an offset term for each systemic velocity from
the different instrumental setups. 
We fitted for  
the scaled orbital distance ($a/R_\star$), 
 the eccentricity ($e$),   the  argument of periastron ($\omega$),   
  the impact parameter \mbox{($b = a \cos (i)/R_\star   \frac{1-e^2}{1+ e sin (\omega)}  $}), 
  the Doppler semi-amplitude variation ($K$), the orbital period ($P_\mathrm{orb}$),  
  the mid-transit time ($T_0$), and the 
 BD-to-star radius  ratio (\rplanet/\rstar). 
We used flat uniform priors over the ranges listed 
 in Table.~\ref{Table: Orbital and planetary parameters},  except for the limb darkening coefficients (LDCs). 
Since the observational cadence of \emph{K2} is close to an integer fraction of the orbital period, 
the data points  appear in clumps in the folded light curve in phase space, as shown in Fig.~\ref{Figure: pyaneti transit model}.    
The ingress and egress are  not well sampled and the LDCs are poorly constrained by the data.
We therefore used Gaussian priors and 
the  \citet{Mandel2002} quadratic limb darkening equation based on the linear 
and quadratic  coefficients $u_1$ and $u_2$, respectively. 
We  used the \citet{Kipping2013} parametrisation \mbox{$q_1 = (u_1 + u_2)^2$} and \mbox{$q_2 = 0.5 u_1(u_1+u_2)^{-1}$}, 
and an interpolation\footnote{{\url{http://astroutils.astronomy.ohio-state.edu/exofast/limbdark.shtml}}  \citep{Eastman2013}}  of the \citet{Claret2011} limb 
darkening tables to our   spectroscopic parameters and the \emph{Kepler} bandpass  
 to set Gaussian priors to $q_1$ and $q_2$.  We used conservative 0.1 error bars
on both the linear and quadratic coefficients.

To account for  the long \emph{K2}  integration time of almost 30~minutes,  
we integrated the transit models  over ten steps  \citep{Kipping2010}.
The parameter space was explored  with 500 independent chains  randomly created inside the prior ranges.
Convergence was checked after every 5\,000 iterations and when   reached, the last
5\,000 iterations were used to create a posterior distribution of 250\,000 independent points for every parameter. 
We removed one outlier from the light curve. Since  $\chi^2/\mathrm{d.o.f}=1.3$, we  
fitted for an RV jitter term  for each instrument setup in the model to take into account  
additional instrumental noise not included in the   uncertainties and    stellar activity-induced  variation, 
and   a light curve jitter term to account for the dispersion of the in- and out-of-transit data to 
obtain $\chi^2/\mathrm{d.o.f}=1.0$.

The high RV amplitude of about 5~\kms~in Fig.~\ref{Figure: pyaneti rv model} immediately signalled that the 
mass of the transiting object is
much higher than the expected mass from a Jupiter-like planet. 
This is not possible to derive from the light curve
alone since BDs and Jupiters have approximately the same size. The final mass is about 5~\% of the stellar host mass. 
We also note  that the BD is near grazing as the derived impact parameter is \bb~which suggests that
the derivation of limb darkening may be less accurate \citep{2013A&A...549A...9C}. 
If \targetaa~had a typical radius of an \stype~star instead 
of being slightly evolved (with about 8~\% larger radius), 
the BD would   be grazing.

\citet{2019arXiv190303118C}  
used    the TRES spectrograph at the  1.5~m Tillinghast telescope at Mt.~Hopkins, Arizona 
with a spectral resolution $R = 44\,000$ covering $390 - 910$~nm  to measure 14 RVs with S/N $\approx 22-45$ of \targetaa. 
This can be compared to our 23 RVs with S/N $\approx 35-75$. 
Their uncertainties are somewhat larger than ours, but our results agree within 1~$\sigma$.

The final results are listed in Table~\ref{Table: Orbital and planetary parameters}. 
We used the median and 68.3~\% credible interval of the
posterior distributions which all were  smooth and unimodal.  
 We show the    folded light curve with the best-fitted transit model in Fig.~\ref{Figure: pyaneti transit model},    
 and the phase-folded RV curve with our best-fitted model in  Fig.~\ref{Figure: pyaneti rv model}.


\begin{table*}
\centering
\caption{Priors to the {\tt{pyaneti}} model of \targetaa b   and results.}
\begin{tabular}{lcrr}
\hline
\hline
\noalign{\smallskip}
Parameter & Units & Priors$\tablefootmark{a}$ & Final value  \\
\noalign{\smallskip}
\hline
\noalign{\smallskip}
              
\multicolumn{3}{l}{\emph{Fitted parameters}}\\
\noalign{\smallskip}                
                ~~~$T_0$ &Transit epoch (\bjdtdb - 2\,450\,000)\dotfill &  $\mathcal{U}$[8098.665,  8098.695]   & \Tzerob             \\
\noalign{\smallskip}                
                ~~~$P_\mathrm{orb}$ &Orbital period (days)\dotfill &  $\mathcal{U}$[5.1679, 5.1719]  &\Pb \\
\noalign{\smallskip}
                ~~~$e$   &Eccentricity \dotfill & $\mathcal{U}$[0, 0.3]   &\eb     \\
\noalign{\smallskip}
                ~~~$\omega$   &Argument of periastron (degrees)\dotfill & $\mathcal{U}$[0,  180]   & \wb     \\                
\noalign{\smallskip}                                
                ~~~$b$  &Impact parameter\dotfill &  $\mathcal{U}$[0, 1]   & \bb  \\
\noalign{\smallskip}                
                ~~~$a/R_\star$  & Scaled semi-major axis\dotfill &  $\mathcal{U}$[1.1, 15]   &\arb   \\
 \noalign{\smallskip}                                      
                ~~~$R_{\mathrm{BD}}/R_\star$ & Scaled  brown dwarf radius\dotfill & $\mathcal{U}$[0, 0.1]    &\rrb \\
\noalign{\smallskip}
                ~~~$K $ & Doppler semi-amplitude variation (\kms)\dotfill &   $\mathcal{U}$[0,15]     &\kb     \\
 
\noalign{\smallskip}                
                ~~~$q_1$ &Parameterised   limb-darkening coefficient\dotfill &  $\mathcal{G}$[0.38, 0.10]   & \qone    \\
\noalign{\smallskip}                              
                ~~~$q_2$  &Parameterised   limb-darkening coefficient\dotfill &  $\mathcal{G}$[0.26, 0.10]   & \qtwo        \\

\noalign{\smallskip}
\multicolumn{3}{l}{\emph{Derived Parameters}}\\               
\noalign{\smallskip}                                             
                  ~~~$M_\mathrm{BD}$ & Brown dwarf mass (\mjup)\dotfill   & \dots     & \mpb  \\
\noalign{\smallskip}                
                   ~~~$R_\mathrm{BD}$ & Brown dwarf radius (\rjup)\dotfill &  \dots    & \rpb \\
\noalign{\smallskip}
                ~~~$i\, \,\tablefootmark{b}$  &Inclination (degrees)\dotfill &   \dots   & \ib  \\                

 \noalign{\smallskip}                                              
                  ~~~$a$ &Semi-major axis (AU)\dotfill &  \dots   & \ab   \\                
 \noalign{\smallskip}                
           ~~~$F$ &Insolation  ($F_\mathrm{\oplus}$)\dotfill & \dots     &\Fequib \\
 \noalign{\smallskip}                
           ~~~$\rho_\star\,\tablefootmark{c}$ &Stellar density  (g~cm$^{-3}$)\dotfill & \dots   & \denspyaneti  \\

 \noalign{\smallskip}                
           ~~~$\rho_\mathrm{BD}$ & Brown dwarf  density  (g~cm$^{-3}$)\dotfill &  \dots   & \denpb \\

 \noalign{\smallskip}                
           ~~~$\log (g_\mathrm{BD})$ & Brown dwarf   surface gravity  (cgs) \dotfill & \dots    & \pgrav \\

\noalign{\smallskip}                
           ~~~$T_{eq}\,\tablefootmark{d}$ &Equilibrium temperature (K)\dotfill & \dots    & \Tequib \\
               
\noalign{\smallskip}                          
                ~~~$T_{14}$ &Total transit duration (hours)\dotfill &  \dots   & \ttotb     \\
 
 \noalign{\smallskip}                          
                ~~~$T_{23}$ &Full   transit  duration (hours)\dotfill & \dots    & \tfullb    \\

\noalign{\smallskip}                
                ~~~$u_1$ &Linear limb-darkening coefficient\dotfill & \dots & \uone    \\
\noalign{\smallskip}                              
                ~~~$u_2$  &Quadratic limb-darkening coefficient\dotfill &   \dots   & \utwo        \\

\noalign{\smallskip}
\multicolumn{3}{l}{\emph{Additional Parameters}}\\      

\noalign{\smallskip}                              
                ~~~$\gamma_1$  &Systemic velocity FIES1 (\kms)\dotfill &  $\mathcal{U}$[-27.1655, -16.5310]    & \fone        \\
\noalign{\smallskip}                              
                ~~~$\gamma_2$  &Systemic velocity FIES2 (\kms)\dotfill &$\mathcal{U}$[-22.9664, -16.9289]     & \ftwo        \\                
 \noalign{\smallskip}                              
                ~~~$\gamma_3$  &Systemic velocity FIES3 (\kms)\dotfill &  $\mathcal{U}$[-27.3661, -16.5940]   & \fthree        \\     
                
\noalign{\smallskip}                              
                ~~~$\sigma_{F1}$ &RV jitter FIES1 (\kms)\dotfill &$\mathcal{U}$[0, 1] &\gammaone       \\        
 \noalign{\smallskip}                              
                ~~~$\sigma_{F2}$ &RV jitter FIES2 (\kms)\dotfill &$\mathcal{U}$[0, 1]   &\gammatwo      \\                    
 \noalign{\smallskip}                              
                ~~~$\sigma_{F3}$ &RV jitter FIES3 (\kms)\dotfill & $\mathcal{U}$[0, 1]  &\gammathree       \\    
  \noalign{\smallskip}                              
                ~~~$\sigma_{tr}$ &Light curve jitter  \dotfill & $\mathcal{U}$[0, 0.00004733] &  \trjitter     \\

\noalign{\smallskip}                
\hline
\end{tabular}
\label{Table: Orbital and planetary parameters}
\tablefoot{
\tablefoottext{a}{$\mathcal{U}$[a,b] refers to uniform priors in the range  \emph{a} --  \emph{b}, and $\mathcal{G}$[a,b] 
refers to Gaussian priors with mean \emph{a} and standard deviation  \emph{b}.} 
\tablefoottext{b}{Orbit inclination   relative to the   plane of the sky.} 
\tablefoottext{c}{Density from pyaneti transit modelling. Density   from adopted stellar mass and
radius is \mbox{\densMR~g~cm$^{-3}$}.} 
\tablefoottext{d}{Assuming isotropic re-radiation and a Bond albedo of zero. Increasing the albedo to e.g. 0,3 and 0.6,
 we find $T_{eq} \approx 1310$ and 1140~K, respectively.}
}
\end{table*}


\section{Discussion} \label{Section: Discussion}

\targetaa~is a rare type of object in the  BD desert. 
In this section we will investigate its formation and tidal circularisation in addition to a comparison of  GPs and BDs in
the mass-density diagram.

 \subsection{Formation}
 There are several different paths to form BDs \citep[for a summary see e.g.][]{2018haex.bookE..95W}. 
Objects all the way from stellar masses down to about 3~\mjup~can form through gravitational collapse 
and turbulent fragmentation like stars \citep{2004ApJ...617..559P, 2008ApJ...684..395H}.  
In   protoplanetary discs, BDs can also form  up to possibly  a few tens of \mjup~according to the 
core-accretion planet formation theory in 
either its traditional planetesimal accretion or later pebble accretion variants 
\citep[e.g.][]{1996Icar..124...62P, 2003ApJ...598L..55R, 2004A&A...417L..25A, 2012A&A...544A..32L, 2012A&A...547A.112M}.  
For  \targetaa b with a mass of \mpb~\mjup, too massive for formation by core accretion, formation by gravitational 
instability in the protoplanetary disc may instead be   possible 
\citep{1964ApJ...139.1217T, 2016ARA&A..54..271K}. 
Disc fragmentation typically occurs at radii 
$>$10~AU and forms fragments with initial masses of a few to a few tens of Jupiter masses 
\citep[see reviews by][]{2016ARA&A..54..271K,2017PASA...34....2N}. 
We show in Appendix~\ref{Section: Appendix B} and Fig.~\ref{Figure: Toomre instabilities} that gravitational instability 
can indeed give rise to fragments with the mass of \targetaa~b. One of these fragments 
must then migrate to the present orbit of \targetaa~b, which can happen 
through Type~I migration \citep{2011MNRAS.416.1971B,2015ApJ...802...56M}, 
although the extent of this is debated in the literature 
\citep{2015ApJ...810L..11S,2018A&A...618A...7V}. On the other hand,  
gravitational instability often gives rise to more than one fragment, 
and in this case the dynamical interactions between fragments enhance 
their migration rate through the disc \citep{2018MNRAS.474.5036F}. Indeed, 
the moderate eccentricity of \targetaa b may be a relic of these 
dynamical interactions, after some reduction by tidal forces.

\begin{figure}[!t]
\centering
\resizebox{\hsize}{!}{
 \includegraphics{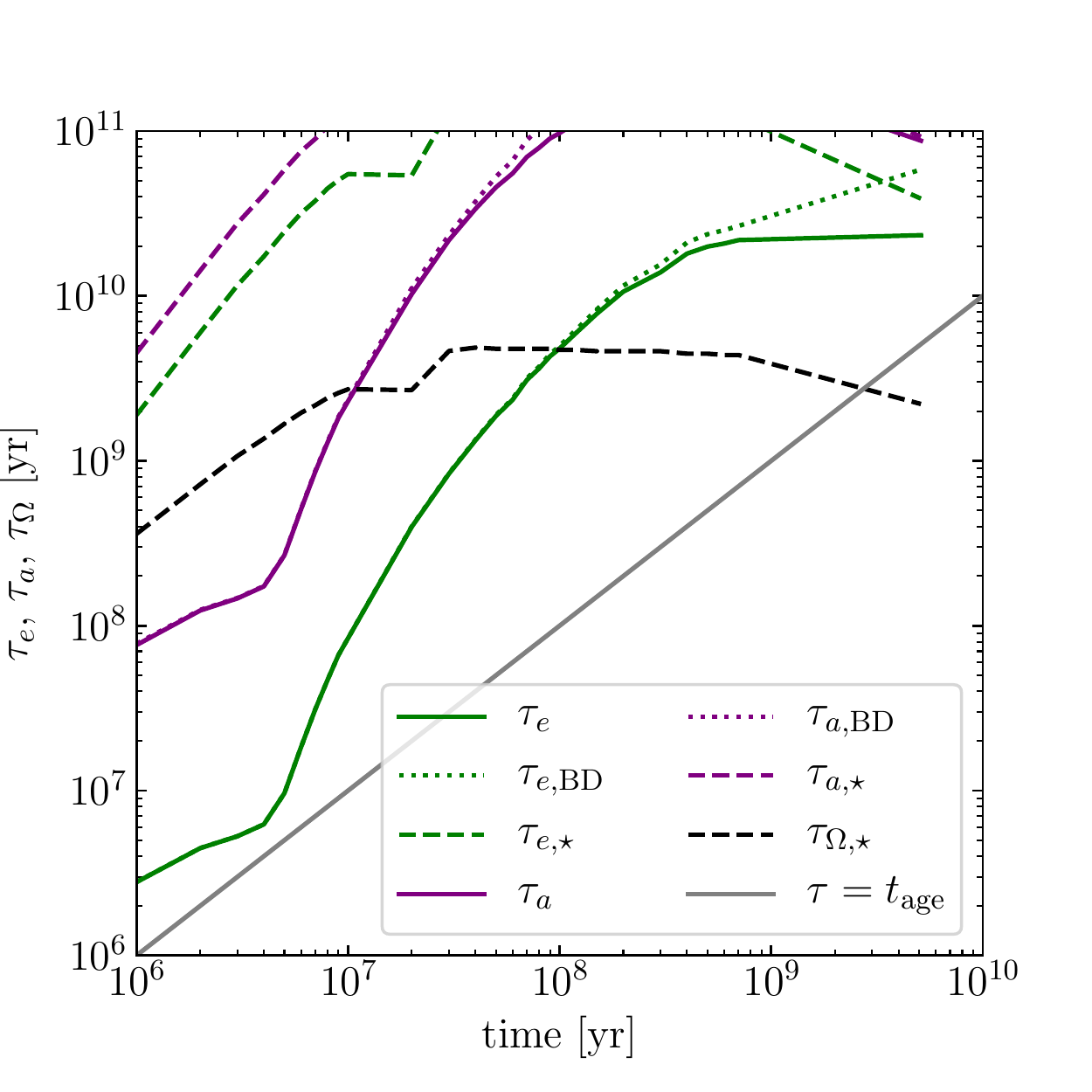}}
 \caption{Timescales for tidal orbital circularisation $\tau_e$, 
semimajor axis decay $\tau_a$, and stellar spin evolution, 
$\tau_{\Omega,\star}$. The contributions from the star and
the brown dwarf as given by \mbox{Eqs.~\ref{eq:tides a_star} -- \ref{eq:tides e_b}}  are shown, as well as 
the net effect, taking quality factors $Q^\prime_\star=10^8$ and  
$Q^\prime_\mathrm{BD}=10^5$. The diagonal line marks where the timescale at a
given age is equal to the system's age.}
    \label{Figure: tidal timescales}
\end{figure}

\subsection{Tidal evolution of the system} \label{subsection: Tidal evolution}

As the BD is on a close orbit with non-zero eccentricity, its orbit may be affected by tidal torques.
These arise either from the deformation of the BD 
by the star (henceforth the planetary tide) or from the deformation of the star 
by the BD (henceforth the stellar tide). 
These tides cause a change in both orbital semi-major axis and eccentricity, 
and hence there are four timescales to consider: the contributions of each tide to 
the decay of the semi-major axis and to the eccentricity. 
We use the tidal model of \citet{2008ApJ...678.1396J} and define the 
following timescales $\tau$:
\begin{eqnarray}
\frac{1}{\tau_{a,\star}} &=& a_\mathrm{BD}^{-13/2}\frac{9}{2}\sqrt{\frac{G}{M_\star}}\frac{R_\star^5 M_\mathrm{BD}}{Q^\prime_\star}   \label{eq:tides a_star} \\
\frac{1}{\tau_{a,\mathrm{BD}}}& = & a_\mathrm{BD}^{-13/2} \frac{63}{2}  \sqrt{GM_\star^3} \frac{R_\mathrm{BD}^5e_\mathrm{BD}^2}{Q^\prime_\mathrm{BD}M_\mathrm{BD}} \label{eq:tides a_b}  \\
\frac{1}{\tau_{e,\star}} &=& a_\mathrm{BD}^{-13/2} \frac{171}{16} \sqrt{\frac{G}{M_\star}} \frac{R_\star^5 M_\mathrm{BD}}{Q^\prime_\star}  \label{eq:tides e_star}  \\ 
\frac{1}{\tau_{e,\mathrm{BD}}}&=&a_\mathrm{BD}^{-13/2}\frac{63}{4}\sqrt{GM_\star^3}\frac{R_\mathrm{BD}^5}{Q^\prime_\mathrm{BD}M_\mathrm{BD}}\ ,  \label{eq:tides e_b}
\end{eqnarray}
where $Q^\prime_\star$ and $Q^\prime_\mathrm{BD}$ are the tidal quality factors of the star and the BD. 
We adopt quality factors of $10^8$ for the star, in line both with the recent empirical calibration of 
\citet{2018MNRAS.476.2542C} for stars in the equilibrium tide regime
and with dynamical tide calculations for a $1.2\mathrm{\,M}_\odot$ F-type 
star by \cite{2007ApJ...661.1180O}, 
and $10^5$ for the BD  as inferred for Jupiter  \citep{2009Natur.459..957L}. 
For simplicity, we hold $Q$ constant for both the star and the brown dwarf. 
In reality, $Q$ can exhibit a complicated dependence on the ratio of the 
periods of the orbit and of the stellar spin: see Fig.~8 of \cite{2009MNRAS.395.2268B}. 
We find that, with the current system parameters, the stellar tide dominates, and the decay 
timescales are $\tau_a = 87$\,Gyr and $\tau_e=23$\,Gyr. These values are longer than the system age, and 
hence the BD's orbit will not be currently tidally evolving.  We note that 
the preprint of
\citet{2019arXiv190303118C} gives a slightly longer circularisation time of 47 Gyr. The 
difference is largely due to them considering only the tide raised on the brown 
dwarf.

Note that the tidal timescales given in Eqs.~\ref{eq:tides a_star} 
-- \ref{eq:tides e_b} are extremely strong functions
of the physical radii of the BD and of the star, so the tidal timescales change with system age 
\citep[see, e.g.,][]{1989A&A...223..112Z,2015A&A...580L...3M,2016CeMDA.126..275B}. 
To explore the historical evolution 
of the tidal forces, we used the PHOENIX BT-Settl models \citep{2015A&A...577A..42B} 
to obtain the radii of both the primary and the BD, and calculated the tidal timescales 
as a function of system age  (see Fig.\ref{Figure: tidal timescales}).
This shows that for the system's main sequence lifetime the tidal forces have been negligible, 
but that the circularisation timescale was comparable to the system age at ages of 
a few Myr, when the BD radius was several $\mathrm{\,R_J}$. 
Thus, it is possible that \targetaa~b started tidally circularising 
early in its history and then stopped as its radius contracted.

A further issue relates to the evolution of the stellar spin: around 98\% of the 
system's angular momentum lies in the brown dwarf's orbit, so it should spin the star up to 
\mbox{(pseudo-)}synchronisation\footnote{Pseudo-synchronisation occurs for eccentric orbits where the spin 
angular velocity locks to a value given by Eq.~(42) of \citet{1981A&A....99..126H}. The exact value is a
function of eccentricity and orbital frequency.}  
if the timescale is short enough. 
For present parameters, pseudo-synchronisation occurs at 
\mbox{$\Omega_\mathrm{rot, ps} = 1.083 \pm 0.003 \Omega_\mathrm{orb}$}, far from 
the actual value of \mbox{($\Omega_\mathrm{rot, actual} / \Omega_\mathrm{orb} = 0.69$)}. 
With \mbox{$Q^\prime_\star=10^8$} we 
find a timescale for spin evolution of \mbox{$\tau_{\Omega,\star}=2.2$\,Gyr}, comparable to the system age.
Given that the star is not pseudo-synchronised, this implies that \mbox{$Q^\prime_\star\gtrsim 10^8$}. 
In principle, $Q^\prime_\star$ can be determined by  
transit timing variations, but this is challenging: from Eq.~7 of 
\cite{2014MNRAS.440.1470B}, we estimate that transits would occur just 1~s 
earlier after 20 years even if $Q^\prime_\star=10^7$. 
Alternatively, magnetic effects such as magnetic breaking may force the system 
away from pseudo-synchronisation: magnetic fields are possessed by 
both BDs (of kG or stronger: 
\citealt{2018ApJS..237...25K, 2017ApJ...847...61B, 2017MNRAS.465.1995M})
and F stars \citep[e.g.][]{2014A&A...562A.124M, Augustson_2013}. 
The stellar wind and the magnetism of the BD,
studied e.g. in \citet{2015ApJ...807...78F}, can also interplay, as well as induction heating \citep{2018ApJ...858..105K}.

We  summarise a potential formation and evolution history for this system:  
\targetaa b formed through gravitational instability early in the protoplanetary 
disc's evolution. It may have formed as one of several similar objects, the others 
either ejected by dynamical interactions or undetectable given current data.
The interactions with the other objects would have excited \targetaa b's orbital eccentricity, 
and hastened its migration towards the primary star in the few Myr of the protoplanetary 
disc's lifetime. At this young age, 
the BD's large radius may have led to some tidal decay of its orbital eccentricity, but 
after several Myr its radius would have shrunk enough to weaken tidal forces enough to 
freeze its orbit in place. Finally, the tide raised on the star by the BD 
may have begun forcing the star towards spin-orbit pseudo-synchronisation during 
the star's main-sequence lifetime, but this process has not yet finished.

 \begin{figure}[!t]
 \centering
 \resizebox{\hsize}{!}{
   \includegraphics{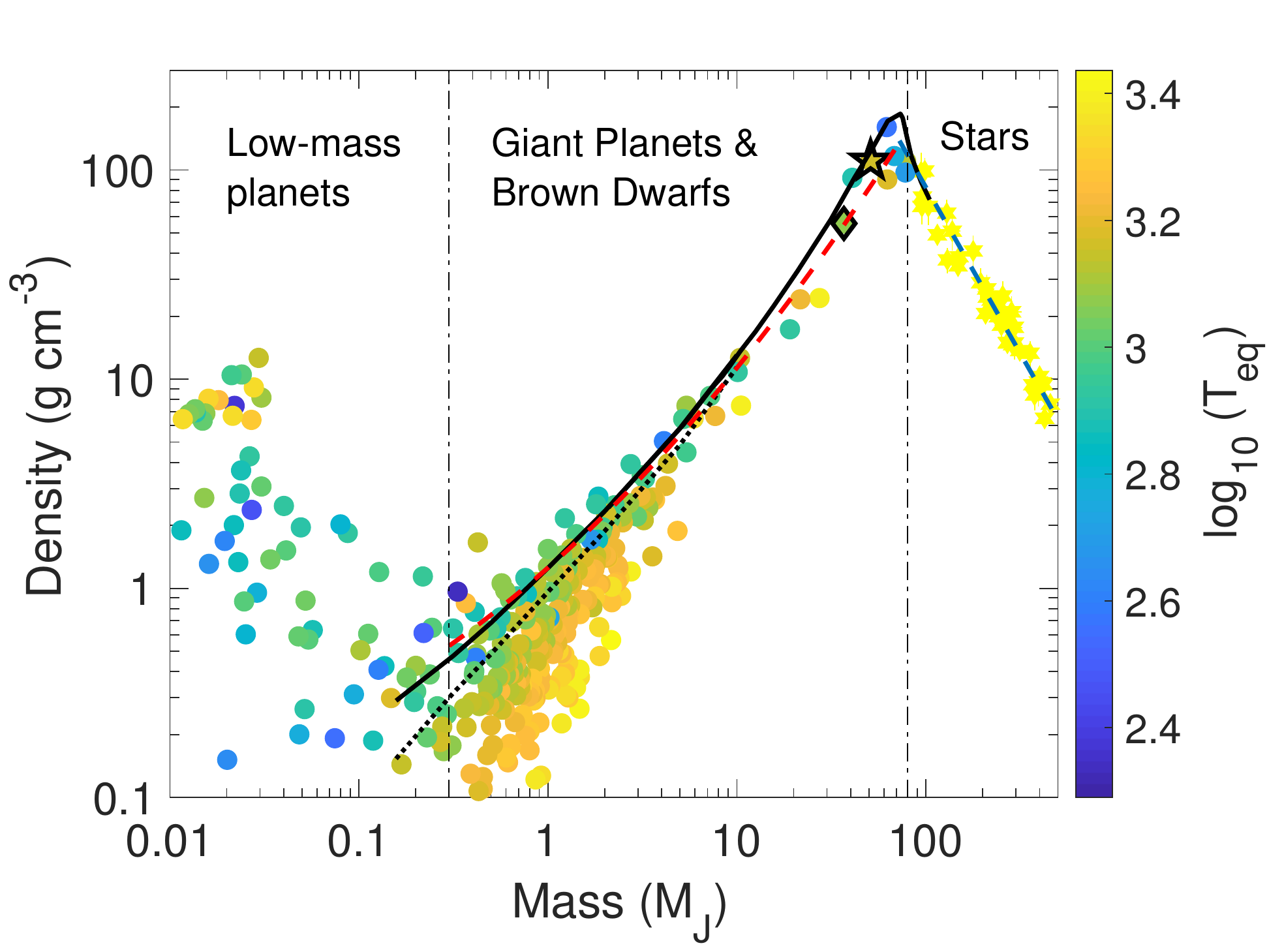}}
   \caption{The mass-density diagram for  planets, brown dwarfs,  
   and    low-mass  stars in eclipsing
binaries with a precision in measured mass and density $<20 \%$. 
The  star and diamond symbols  mark the locations  
of  \targetaa b and EPIC~219388192b also found by our programme 
\citep{2017AJ....153..131N}.  
The red dashed line  represent    a second order polynomial fit to the data  with \mbox{$M = 0.3 - 80$~\mjup}~
and equilibrium temperatures \mbox{$<1000$~K}. The blue dashed line shows a linear fit to the stars with \mbox{$M > 80$~\mjup}. 
The nominal separation at 80~\mjup~between brown dwarfs and stars, and the empirical separation between low-mass and giant planets
at 0.3~\mjup, are marked with the vertical dashed-dotted lines. 
The solid   black line   shows the theoretical relationship for H/He dominated giant objects  with  \mbox{$Z = 0.02$}, 
age = 5~Gyr, without irradiation \citep{2003A&A...402..701B, 2008A&A...482..315B}, and the dotted black line the same model including 
irradiation from a solar-type star at 0.045~AU  \citep{2008A&A...482..315B}.   
}
      \label{Figure: mass-density diagram}
 \end{figure}

\subsection{Mass-density diagram} \label{subsection: Mass-density diagram}

In order to investigate possible differences between BDs and GPs,  
 we  show   a mass-density diagram in Fig.~\ref{Figure: mass-density diagram}   
for     planets\footnote{Well-studied planets listed at \url{http://www.astro.keele.ac.uk/jkt/tepcat/}.}
and 
 BDs\footnote{References in Sect.~\ref{Section: introduction}.  
discovered by space- and ground-based transit searches.}.  
It should be noted that all these objects have close-in orbits to their host star (most have \mbox{$P_\mathrm{orb} < 10$~days}). 
 Also shown are eclipsing low-mass 
stars\footnote{\citet{2003A&A...398..239R, 2005A&A...431.1105B, 2005A&A...438.1123P, 2006A&A...447.1035P, 2009A&A...505..205D, 2013A&A...553A..30T, 2014MNRAS.437.2831Z, 2014A&A...572A.109D}; 
\citet{2016MNRAS.462..554C} and references in Table~1; 
 \citet{2017ApJ...849...11G, 2017A&A...604L...6V, 2017ApJ...847L..18S}; 
 \citet{2018AJ....156...27C}  and references in Table~4; \citet{2019arXiv190303118C}.} 
up to 450~\mjup~(0.43~\Msun)  
mostly from ground-based discoveries. 
We only include objects   with a precision in mass and density better than 20~\% (in total 253 GPs and BDs, and 43 low-mass stars). 
The vertical dashed-dotted line at 80~\mjup~marks the  nominal   separation between BDs and nuclear burning M dwarfs. 
The colours of the planets and brown dwarfs indicate the logarithm of the  equilibrium temperatures, $T_\mathrm{eq}$. 
It is clearly seen that low-mass GPs with high incident flux,  
and thus high  $T_\mathrm{eq}$, have lower densities which could be a sign of inflated radii due to the proximity to the host stars 
\citep{2011ApJ...736L..29M, 2014prpl.conf..763B, 2017ApJ...841...30T}. 
We fitted a second order polynomial to the data (red dashed line) 
between 0.3~\mjup~and 80~\mjup~for objects  with 
\mbox{$T_\mathrm{eq} <1000$~K}   to exclude  objects with inflated radii  \citep{2013ApJ...768...14W}, in total 33 objects, and found 
\mbox{$\log \rho = 0.16 \times \log^2 (M) + 0.80\times \log (M) + 0.10$}. The blue dashed line shows a linear fit to the stars with \mbox{$M > 80$~\mjup}: 
\mbox{$\log (\rho) = -1.6\times \log (M) + 5.1$}.  
Compared to \citet{2015ApJ...810L..25H},  
we now find a sharp turn-over at  $\sim73$~\mjup~instead
of   $\sim60$~\mjup. 
The empirical fit follows closely the 
 theoretical relationship for H/He dominated {GPs \citep{2008A&A...482..315B} and BDs  \citep{2003A&A...402..701B} with  \mbox{$Z = 0.02$}, 
\mbox{age = 5~Gyr} and without irradiation drawn with a solid   black line.  
The dotted black line shows the same model including 
irradiation from a solar-type star  at $a = 0.045$~AU  \citep{2008A&A...482..315B} which clearly shows the impact of irradiation for the lower mass
GPs.    
At the lower end, we find  a  turn-over at $\sim 0.3$~\mjup~in agreement with  \citet{2015ApJ...810L..25H}, marking the transition to low-mass planets. 
Our results are in agreement with  \citet{2017ApJ...834...17C} who 
found $R \sim M^{-0.04}$  for objects between  0.4~\mjup~and 80~\mjup.

Our two BDs fall close to the theoretical model for H/He dominated BDs  as 
well as   the empirical fit.   
 No   distinguishing features 
between   GPs and BDs can be seen.  
After a brief phase  lasting $\sim$10~Myr when the deuterium and lithium fusion    halts contraction, 
BDs cool and contract in a way similar to GPs.  
This suggests that both types of objects   will follow   the same trend in the mass-density diagram
independent of the formation mechanism, especially at late ages. 
At earlier stages, the difference in radius is   larger  \citep[e.g.][]{2008A&A...482..315B}
and contributes to the scatter of the data points.

\section{Conclusions}\label{Section: Conclusions}
We report the discovery and characterisation of  a rare object with a mass of \mpb~\mjup~and
a radius of   \rpb~\rjup~in an eccentric 5.17~day   orbit around 
    the slightly evolved \stype~star \targetaa. 
Since the star is seen close to equator-on, future observations 
with large  8--10~m class telescopes,
could   allow the measurement of the (3-D) obliquity  
 angle between the stellar rotation axis and the 
brown dwarf  orbit axis via the Rossiter-McLaughlin effect. 
Thanks to the evolutionary state of the host star, this is one of the few transiting  brown dwarfs for which 
a relatively precise age can be estimated. 
    Our results are in agreement   with \citet{2019arXiv190303118C} who recently 
reported an independent
discovery  and characterisation of \targetaa b.

 We show with a simple analytical model that formation of a brown dwarf of the required mass 
is possible at several tens of AU through gravitational instability, although significant 
orbital migration is required to bring the object to its current orbit. 
 The orbit may have experienced  a period  of tidal circularisation 
within the first few Myr of the system's life when the brown dwarf's radius was 
very much larger than it is at present, which ceased as its physical radius 
contracted. The stellar spin may have been affected by 
the tidal torque from the BD during the system's main-sequence lifetime, 
but the lack of spin--orbit synchronisation points to a weak stellar 
dissipation parameter ($Q^\prime_\star\gtrsim10^8$). There is also a possibility that
magnetic field plays a role here which could 
change this estimate.

 We find no distinction between brown dwarfs and giant planets based on the mass-density diagram.
This supports the previous suggestion by \citet{2015ApJ...810L..25H},  and supported by \citet{2017ApJ...834...17C}, that  
BDs could    simply   represent the high mass end of GPs and that there are  no
observable differences between mature    BDs and GPs. 
The BD desert may   be a reflection of the decreasing number of objects towards the
high mass end of the GP distribution formed by core-accretion, and the low-mass end 
of stars formed by gravitational instabilities.

 \begin{acknowledgements}
This paper is based on observations obtained with 
(a) The MuSCAT2 instrument, developed by ABC, at Telescopio Carlos S\'anchez operated on the island of Tenerife by the IAC in the Spanish Observatorio del Teide.
(b) The Nordic Optical Telescope (NOT) operated on 
the island of La Palma jointly by Denmark, Finland, Iceland, Norway, and Sweden, in the Spanish Observatorio 
del Roque de los Muchachos (ORM) of the Instituto de Astrod\'isica de Canarias (IAC; 
CAT and TAC programmes 57-015, 57-206, and 57-210,   OPTICON programmes 2018A-044 
and  2018B-052, and the Spanish-Nordic programme 58-301).  
(c) The McDonald observatory operated by The University of Texas at Austin;
(d) The Subaru Telescope operated by the National Astronomical Observatory of Japan; 
(e) NESSI,  funded by the NASA Exoplanet Exploration Program and the NASA Ames Research Center. 
NESSI was built at the Ames Research Center by Steve B. Howell, Nic Scott, Elliott P. Horch, and Emmett Quigley; 
(f) This paper includes data collected by the \emph{K2} mission. Funding for the \emph{K2} mission is provided by the NASA Science Mission directorate.
We thank the  NOT,  McDonald, Subaru, and NESSI staff members for their  support during the 
observations. 
This work 
has made use of SME package, which benefits from the continuing development work by J. Valenti and N. Piskunov 
and we gratefully acknowledge their continued support. 
\citep{Kupka2000, Ryabchikova2015}.  
C.M.P.  and M.F. gratefully acknowledge the support of the  Swedish National Space Agency (DNR 174/18).  
Sz.Cs. thanks the Hungarian National Research, Development and Innovation Office, for the NKFI-KH-130372 grants.
AJM and MBD acknowledge support from the IMPACT grant from the Knut and Alice Wallenberg Foundation (2014.0017). 
JK, SG, MP, SC, APH, KWFL, ME and HR acknowledge support by DFG grants PA525/18-1, PA525/19-1, PA525/20-1, HA 3279/12-1 and RA 714/14-1within the DFG Schwerpunkt SPP 1992, ``Exploring the Diversity of Extrasolar Planets''.  
G.N., H.J.D and D.N, acknowledge support by grants ESP2015-65712-C5-4-R and ESP2017-87676-C5-4-R of the Spanish Secretary of State for R\&D\&i (MINECO). 
PGB acknowledges support by the MINECO-postDoctoral fellowship prog. ``Juan de la Cierva Incorporacion'' (IJCI-2015-26034). 
PK acknowledges support of GACR 17-01752J.  
MS acknowledges the Postdoc$\textcircled{a}$MUNI project CZ.02.2.69/0.0/0.0/16-027/0008360. 
I.R. acknowledges support from the Spanish Ministry for Science, Innovation and Universities (MCIU) and the Fondo Europeo de Desarrollo Regional (FEDER) through grant ESP2016-80435-C2-1-R, as well as the support of the Generalitat de Catalunya/CERCA programme.
NN acknowledge support  by JSPS KAKENHI Grant Numbers JP18H01265 and 18H05439, and JST PRESTO Grant Number JPMJPR1775. 
LGC acknowledges support from the MINECO FPI-SO doctoral research project SEV-2015-0548-17-2 and predoctoral contract BES-2017-082610.
We thank the anonymous referee whose constructive comments led to an improvement of the paper. 
 \end{acknowledgements}

\bibliographystyle{aa}
\bibliography{references}
 
 \appendix

  \section{Additional Figures and Tables} 
 
\begin{table}[!t]
 \centering
 \caption{FIES  RV measurements of  \targetaa.}
 {\begin{tabular}{lcc}
\hline
\hline
\noalign{\smallskip}
\bjdtdb\tablefootmark{a}&   \multicolumn{2}{c}{RV}      \\
(-2\,450\,000.0) &  (\kms)  &  (\kms)      \\
\noalign{\smallskip}
\hline
\noalign{\smallskip}
FIES~1 & Value & Error \\
 8218.479167 & -27.0655 & 0.0490 \\
 8220.404275 & -18.5041 & 0.0458 \\
 8221.487928 & -16.6310 & 0.0403 \\
 8222.391892 & -19.6053 & 0.0523 \\
 8233.440969 & -24.5166 & 0.0364 \\
 8235.385185 & -21.1938 & 0.0465 \\
\noalign{\smallskip}
FIES~2 \\
\noalign{\smallskip}
 8245.450230 & -22.8664 & 0.0877 \\
 8246.452557 & -17.8474 & 0.0563 \\
 8247.446950 & -17.0289 & 0.0730 \\
\noalign{\smallskip}
FIES~3  \\
\noalign{\smallskip}
 8251.403981 & -18.6691 & 0.0357 \\
 8252.445391 & -16.6940 & 0.0259 \\
 8253.439344 & -19.7521 & 0.0260 \\
 8257.445193 & -16.6979 & 0.0259 \\
 8258.437890 & -18.8870 & 0.0202 \\
 8260.434739 & -26.0322 & 0.0294 \\
 8261.435608 & -20.1255 & 0.0259 \\
 8518.672339 & -26.9764 & 0.0401 \\
 8522.638533 & -22.1673 & 0.0255 \\
 8523.662787 & -27.2661 & 0.0228 \\
 8524.698470 & -22.4489 & 0.0242 \\
 8539.620870 & -25.9885 & 0.0264 \\
 8540.645300 & -19.9479 & 0.0337 \\
 8541.620968 & -16.8566 & 0.0251 \\
\noalign{\smallskip}
\hline
\end{tabular}
}
\tablefoot{
\tablefoottext{a}{Barycentric Julian day in barycentric dynamical time.} 
}
 \label{Table: RV measurements}
\end{table}

\begin{figure}[!t]
\centering
 \resizebox{\hsize}{!}{
   \includegraphics{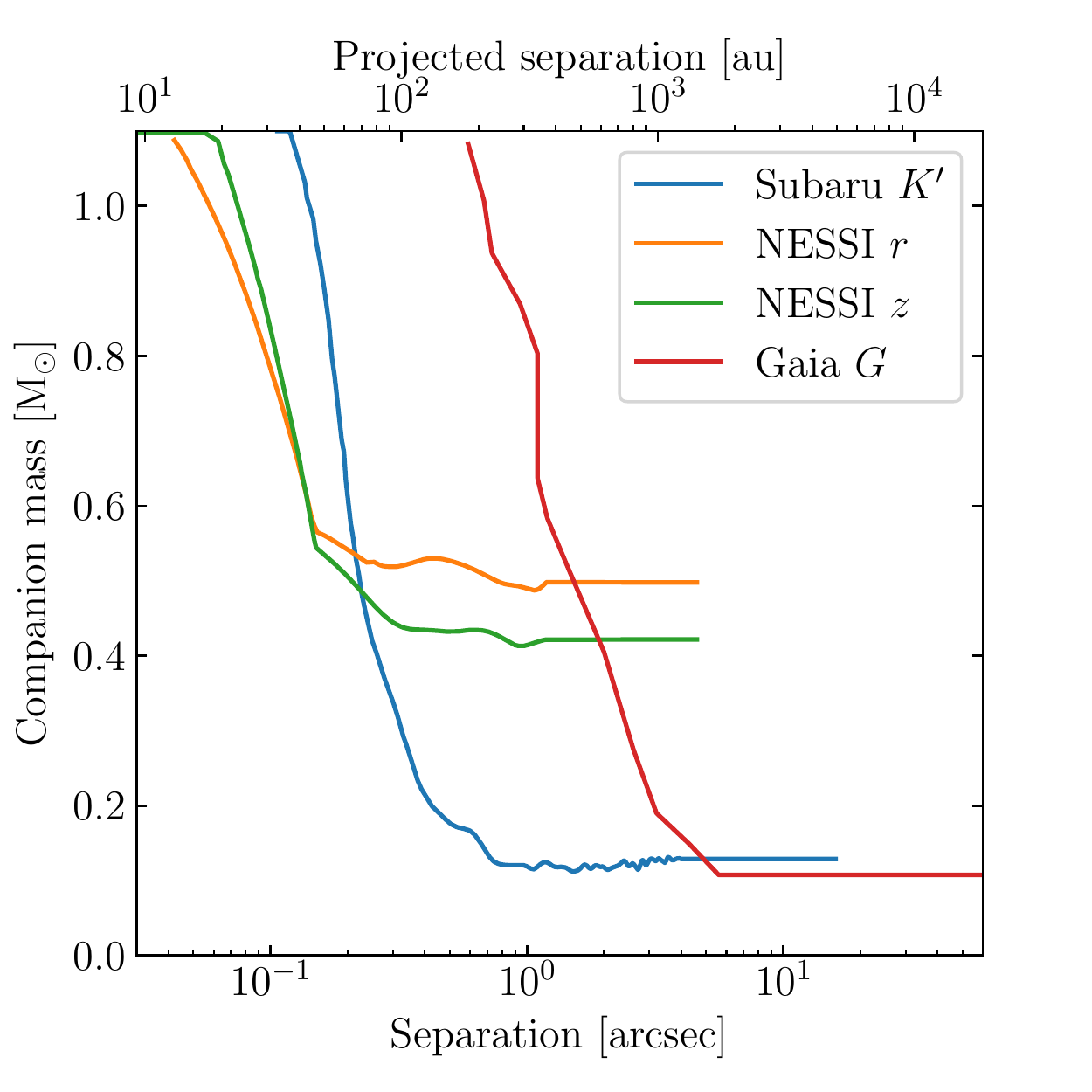}}
  \caption{Limits of companion masses as a function of separation in arcsec and projected separation 
  computed with the \citet{2015A&A...577A..42B} models 
  for our NESSI and Subaru imaging, and the Gaia 50\% detectability limit from \citet{2019A&A...621A..86B}. 
}
     \label{Figure: contrast curve masses}
\end{figure}
 \begin{figure}[!t]
 \centering
  \resizebox{\hsize}{!}{
   \includegraphics[width=\linewidth]{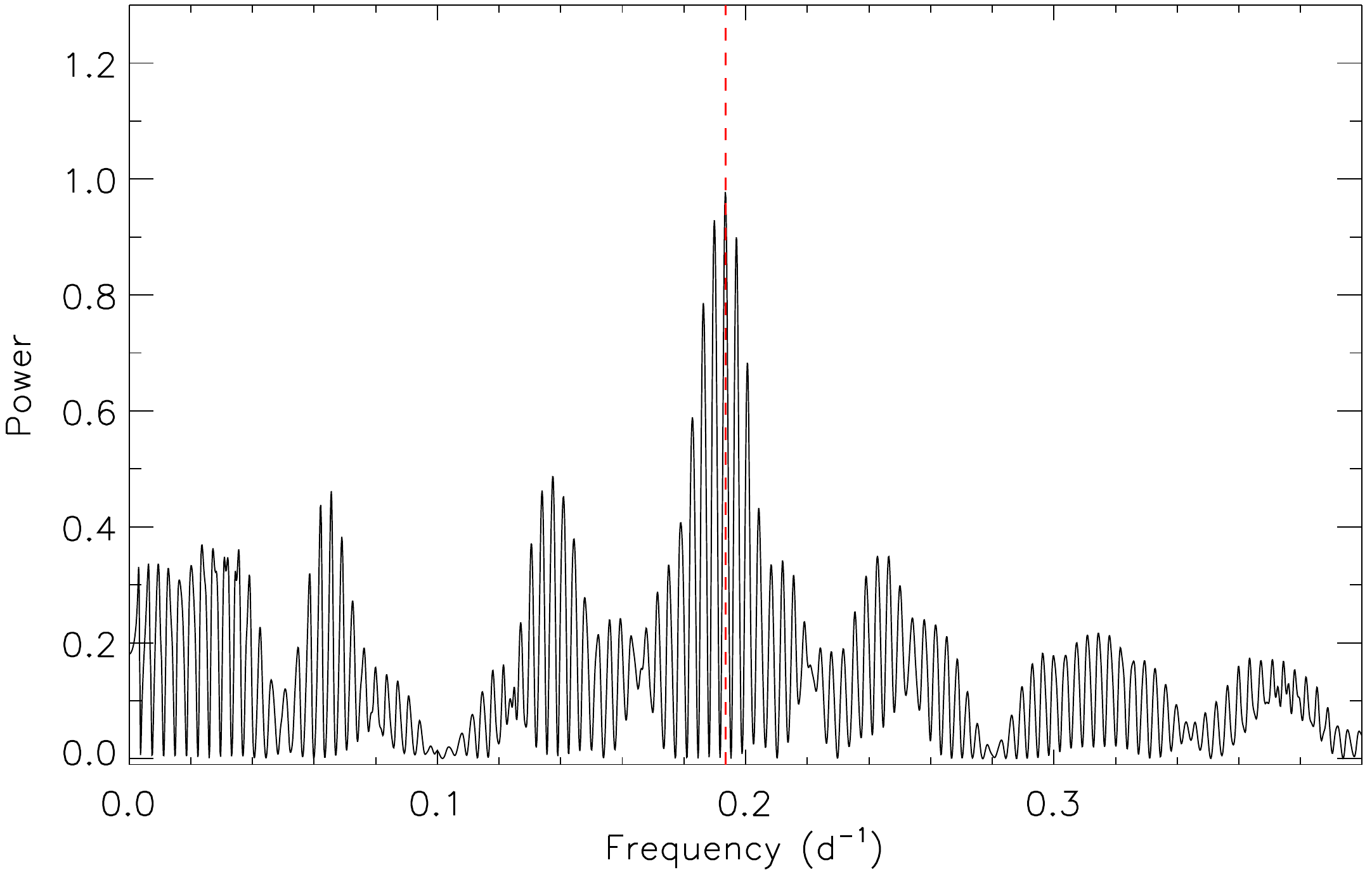}}
   \caption{Generalised Lomb-Scargle periodogram of the combined FIES RVs. The red dashed line marks the orbital frequency of the brown dwarf. Note the presence of the \mbox{1-year} aliases symmetrically distributed around the orbital frequency.}  
      \label{Fig: RV periodogram}
 \end{figure}
 \begin{figure}[!t]
 \centering
  \resizebox{\hsize}{!}{
   \includegraphics[width=\linewidth]{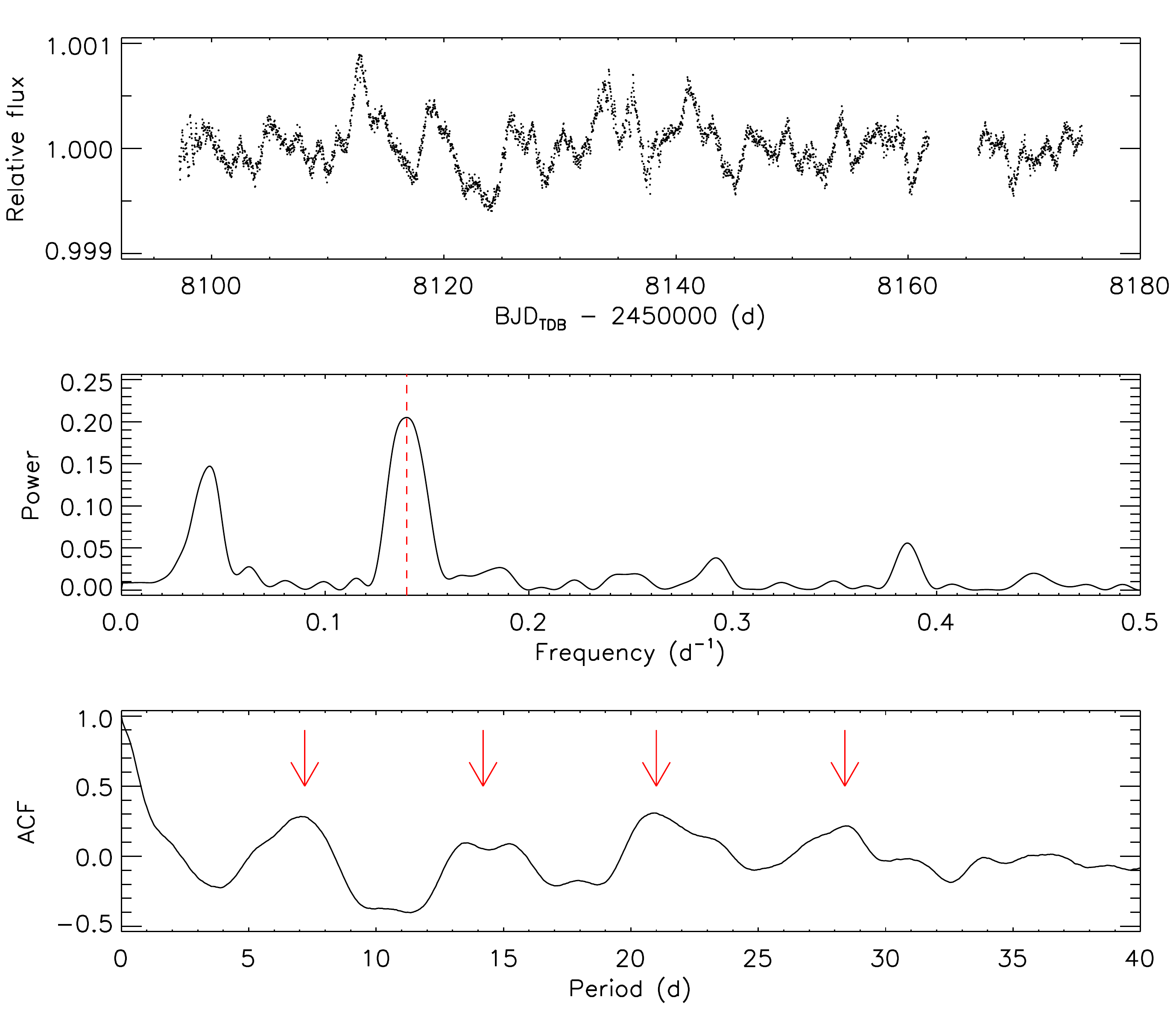}}
   \caption{
\emph{Upper panel}: \emph{K2} light curve of \targetaa~following the removal of 
 the in-transit data-points and the division by the best-fitting 4th-order cubic spline. 
 \emph{Middle panel}: GLS periodogram of the light curve. The red dashed line marks 
 the peaks at the rotation period of the star ($\sim$7 days). \emph{Lower panel}: ACF 
 of the light curve. The red arrows mark the rotation period and its first three harmonics.}  
      \label{Fig: light curve periodogram}
 \end{figure}
%


\section{Formation by gravitational instability}

\label{Section: Appendix B}

Given current uncertainties in both the initial masses of fragments formed by gravitational 
instability, and their subsequent growth and migration \citep{2016ARA&A..54..271K, 2019arXiv190108089F},
we evaluate the prospects for formation by disc instability using simple 
analytical prescriptions. We use the disc model of \citet{2016A&A...591A..72I}, where 
the disc structure is determined by the viscosity, $\alpha$, and the 
mass flux through the disc,  $\dot M_\mathrm{disc}$. We evaluate at which 
radii it is Toomre unstable, and if so, whether the mass of \targetaa b is consistent with   the expected fragment mass 
according to Eq.~49 in \citet{2016ARA&A..54..271K}. The fragment masses are shown in 
Figure~\ref{Figure: Toomre instabilities}. A self-gravitating disc maintains a viscosity 
$\alpha>0.01$, while Class~I YSOs (Young Stellar Objects) and FUORs (FU Orionis stars) 
have mass accretion rates up to a few times $10^{-5}$
\citep{2007ApJS..169..328R, 2014AJ....147..140G}. In these parameter ranges, our model forms 
fragments of several tens of Jupiter masses at $>10$~AU, in agreement with previous works.

\begin{figure}[!t]
\centering
 \resizebox{\hsize}{!}{
   \includegraphics{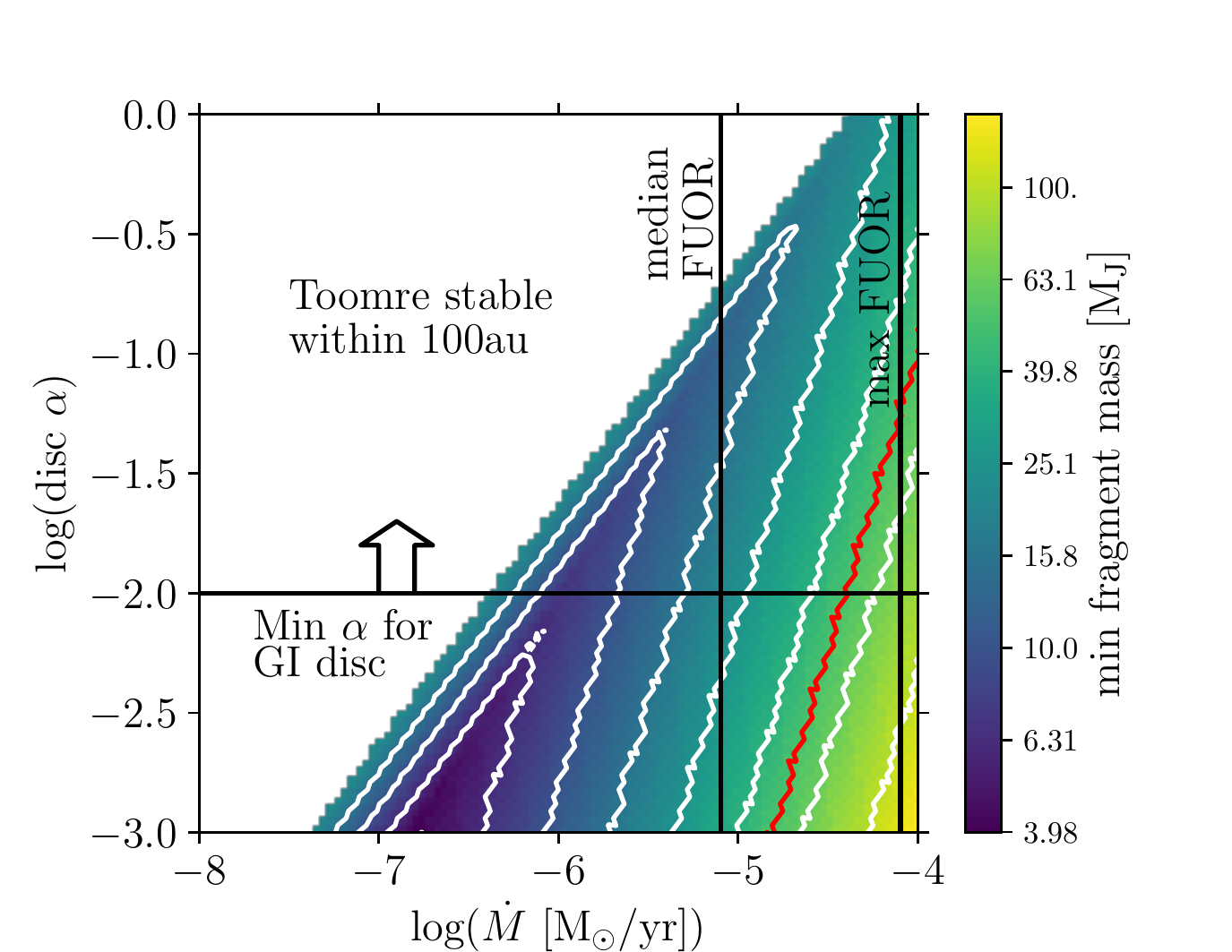}}
  \caption{Formation of \targetaa b by gravitational instability in a 
 protoplanetary disc. The contour plot shows the minimum mass of a 
 fragment arising from disc instability, as a function of the disc's 
 viscosity and accretion rate. 
The red line marks masses equal to the observed mass of \targetaa b. 
The horizontal black line marks the minimum $\alpha$ that a gravitationally 
unstable disc will generate, while the vertical black lines mark the 
median and maximum accretion rates for FUOR discs found by 
\cite{2014AJ....147..140G}. 
 Discs in the white region to the left 
 are gravitationally stable and hence do not form any fragments. 
}
     \label{Figure: Toomre instabilities}
\end{figure}
%
%
 

\end{document}